\documentclass[aps,superscriptaddress, showpacs,preprintnumbers, superscriptaddress, nofootinbibt, twocolumn]{revtex4-2}
\usepackage{eurosym}
\usepackage{amsfonts}
\usepackage{amssymb,amsmath,enumitem}
\usepackage{graphicx,color,epstopdf}
\usepackage{subfigure}
\usepackage{ulem}
\usepackage{tabularx}
\usepackage{booktabs}

\def\be{\begin{equation}}
\def\ee{\end{equation}}
\def\beg{\begin{align}}
\def\eeg{\end{align}}
\def\bea{\begin{eqnarray}}
\def\eea{\end{eqnarray}}

\begin{document}

\title{Dark matter as a Weyl geometric effect}
\author{Piyabut Burikham}
\email{piyabut@gmail.com}
\affiliation{High Energy Physics Theory Group, Department of Physics,
Faculty of Science, Chulalongkorn University, Bangkok 10330, Thailand,}
\author{Tiberiu Harko}
\email{tiberiu.harko@aira.astro.ro}
\affiliation{Department of Theoretical Physics, National Institute of Physics
and Nuclear Engineering (IFIN-HH), Bucharest, 077125 Romania,}
\affiliation{Department of Physics, Babes-Bolyai University, Kogalniceanu Street,
	Cluj-Napoca 400084, Romania,}
\affiliation{Astronomical Observatory, 19 Ciresilor Street,
	Cluj-Napoca 400487, Romania,}
\author{Kulapant Pimsamarn}
\email{fsciklpp@ku.ac.th}
\affiliation{Department of Physics, Faculty of Science, Kasetsart University, Bangkok 10900, Thailand}
\author{Shahab Shahidi}
\email{s.shahidi@du.ac.ir}
\affiliation{School of Physics, Damghan University, Damghan, Iran}

\date{\today }

\begin{abstract}
We investigate the possibility that the observed behavior of test particles outside galaxies, which is usually explained by assuming the existence of dark matter, is the result of the dynamical evolution of particles in a Weyl type geometry, and its associated conformally invariant Weyl geometric quadratic gravity.  As a first step in our investigations we write down the simplest possible conformally invariant gravitational action, constructed in Weyl geometry, and containing the Weyl scalar, and the strength of the Weyl vector only. By introducing an auxiliary scalar field, the theoretical model can be reformulated in the Riemann geometry as scalar-vector-tensor theory, containing a scalar field,  and the Weyl vector, respectively. The field equations of the theory are derived in the metric formalism, in the absence of matter. A specific static, spherically symmetric model, in which the Weyl vector has only a radial component, is considered. In this case, an exact analytic solution of the gravitational field equations can be obtained.  The behavior of the galactic rotation curves is also considered in detail, and it is shown that an effective geometric mass term, with an associated density profile, can also be introduced. Three particular cases, corresponding to some specific functional forms of the Weyl vector, are also investigated. A comparison of the model with a selected sample of galactic rotation curves is also performed when an explicit breaking of conformal invariance is introduced, which allows the fix of the numerical values of the free parameters of the model. Our results show that Weyl geometric models can be considered as a viable theoretical alternative to the dark matter paradigm.
\end{abstract}

\maketitle
\tableofcontents


\section{Introduction}

The development of theoretical physics in the early twentieth century was closely related with the advances in mathematics. The general theory of relativity, proposed by Einstein \cite{Ein}, and Hilbert \cite{Hilb}, is based on the geometry previously proposed by Riemann \cite{Riemm}, and further developed into an elegant and powerful mathematical formalism by Ricci and Levi-Civita \cite{Ric}. The general theory of relativity provides an extremely precise description of the gravitational physics at the level of the Solar System, including such diverse effects as the perihelion precession of Mercury, the light deflection, the Shapiro time delay, the Nordtvedt effect in lunar motion, and frame-dragging, respectively \cite{Will}. The predictions of general relativity have been also brilliantly confirmed by the discovery of the gravitational waves \cite{GW1,GW2}, a discovery that did open a new window on the Universe.

Nevertheless,  recently general relativity began to be confronted  with a number of very serious theoretical challenges. The first major problem did appear from the data resulting from the observations of the Type Ia supernovae \cite{AU1,AU2,AU3,AU4,AU5,AU6,AU7}, indicating that the Universe
is presently in a de Sitter type, accelerating phase.  The data obtained by the Planck satellite \cite{Pl}, and  the studies of the Baryon Acoustic Oscillations \cite{BAO1,BAO2,BAO3} also confirmed the accelerating expansion of the Universe. For a review of the cosmic acceleration problem see \cite{Wein}.

 The simplest theoretical model that could explain the accelerated expansion of the Universe is realized by allowing for the presence of the cosmological constant $\Lambda$ in the gravitational field equations, as suggested a long time ago by Einstein \cite{Einl}.  The inclusion in Einstein's field equations of the cosmological constant gives a very good fit of the observational data. However, this remarkable result is achieved by introducing a new fundamental parameter $\Lambda$, whose physical (or geometrical) nature is not (yet) known. Therefore, in order to find a solution to some of the basic theoretical problems of cosmology, without the need of reintroducing $\Lambda$, the presence of a new fundamental constituent of the Universe, called dark energy, was assumed (see \cite{DE1,DE2,DE3,DE4} for reviews on the dark energy mystery).

Another fundamental problem in present day cosmology and astrophysics is represented by the dark matter problem (for detailed reviews presenting the recent results on the search for dark matter, and for its physical and astrophysical properties see \cite{DM1,DM2,DM3}). The presence of dark matter at galactic and extragalactic scales is necessary for finding a consistent physical explanation of two fundamental astronomical/astrophysical observations, namely, the properties of the galactic rotation curves, and the virial mass deficiency problem in clusters of galaxies, respectively. A large number of astrophysical  observations of the galactic rotation curves \cite{RC1,RC2,RC3,RC4} have conclusively shown that Newton's theory of gravity, as well as general relativity in its standard formulation, cannot properly describe the galactic dynamics of massive test particles orbiting around the galactic center. To explain the observational properties of the galactic rotation curves, and to simultaneously solve the missing mass problem in clusters of galaxies, we need to assume the existence of another dark constituent of the Universe, which is located in a static spherically symmetric halo around the galaxies, and which interacts only gravitationally with ordinary baryonic matter. In  most of the theoretical models dealing with it,  dark matter is described as a cold, pressureless cosmic fluid. There are many particles that have been proposed as dark matter candidates, including axions, neutrinos, WIMPs (Weakly Interacting Massive Particles), neutralinos, gravitinos, neutralinos etc. (see \cite{RDM1,RDM2,RDM3,RDM4} for  reviews of the candidates for the dark matter particle). Dark matter may also be in the form of a Bose-Einstein Condensate \cite{B1,B2,B3}.

However, despite decades of intensive experimental effort, no evidence for the existence of any dark matter particle has been found yet. This situation may require a drastic change our understanding of dark matter, one interesting possibility being that the phenomenology usually explained through the presence of dark matter may in fact be explained as a manifestation of the gravitational force itself. This implies that at large astrophysical and cosmological scales Einstein's general relativity breaks down, and a new theory of gravity may be responsible for both dark matter, as well as for the accelerated expansion of the Universe. Therefore, modified theories of gravity have been extensively used to provide a purely gravitational explanation for dark matter \cite{Mo1, Boh, HMP, HMP1,HMP2, MG1,MG2,MG3,MG4,MG5,MG6,MG7,MG8, MDM1, MDM2,MDM3,MDM4, MDM5,MDM6,MDM7,MDM8, book}. A renormalization group improvement of the Einstein-Hilbert action was considered in \cite{Reuter}, by considering Newton's constant and the cosmological constant to be scalar functions on spacetime.  A power law running of Newton's constant, with a small exponent of the order $10^{-6}$ would account for the non-Keplerian behavior of the rotation curves, without having to assume the presence of any dark matter in the galactic halo.

A few years after the development of the theory of general relativity, essentially  based on Riemannian geometry,  Weyl \cite{Weyl1, Weyl2} did introduce an interesting generalization of Riemann geometry. Weyl's explicit aim was to formulate a unified theory of the electromagnetic and of the gravitational interactions. Since in vacuum  Maxwell's equations of electromagnetism are conformally invariant, Weyl suggested that the field equations of gravity should have the same symmetry. The first important feature of Weyl's geometry is its nonmetric character, since in this geometry the covariant derivative of the metric tensor is nonzero, $\nabla _{\mu}g_{\alpha \beta}=Q_{\mu \alpha \beta}=\omega_{\mu}g_{\alpha \beta}$, where $Q_{\mu \alpha \beta}$ is the nonmetricity, while $\omega _{\mu}$ is called the Weyl vector field. For an introduction to Weyl geometry, its historical development, and its possible physical applications see \cite{Weyl3}. The second important characteristic of Weyl geometry is that the length of a vector varies during the parallel transport. This property of vectors in Weyl geometry is at the basis of Einstein's criticism of the  physical interpretation of the new geometry, as proposed by Weyl. Einstein reasoned that, since the physical properties of the atomic clocks depend on their previous evolution, in an electromagnetic field sharp spectral lines cannot exist.

Nevertheless, if one abandons the assumption of the Weyl vector field as corresponding to an electromagnetic type potential, the conformally invariant Weyl geometry leads to a powerful and beautiful  extension of Riemann's geometry.

 Weyl's theory of  gravity was generalized by Dirac \cite{Di1,Di2}, who extended the theory by introducing a real scalar field $\beta$ of weight $w(\beta)=-1$.  The cosmological applications of the conformally invariant Dirac theory have been studied in \cite{Di3} and \cite{Di4}, respectively. Other geometrical and physical generalizations of the Weyl theory have been considered in \cite{Ut1,Ut2,Ni}.

Conformal invariance of gravitational theories in the framework of Riemann geometry, and its physical implications, has also attracted a lot of attention. A conformally invariant gravitational theory, based on the action
\begin{equation}\label{W1}
S_{Weyl} =-\frac{1}{4}\int d^4x\sqrt{-g}
C_{\mu\nu\rho\sigma}C^{\mu\nu\rho\sigma},
\end{equation}
where $C_{\mu\nu\rho\sigma}$ is {\it the Weyl tensor}, was investigated in detail in \cite{M1,M2,M3,M4,M5,M6}. These types of theories are called {\it conformally invariant, or Weyl gravity type theories}. But note that these theories are not formulated in the proper Weyl geometry, and they have a purely Riemannian geometric basis. In the case of a  static, spherically symmetric, metric of the form
\be
ds^2=-B(r)dt^2+B^{-1}(r)dr^2+r^2d\Omega,
\ee
 the Weyl gravity theory has vacuum solutions of the form
 \be\label{solM}
 B(r)=1-3\beta \gamma -\frac{\beta\left(2-3\beta \gamma\right)}{r}+\gamma r+kr^2,
 \ee
 where $\beta$, $\gamma$ and $k$ are constants \cite{M1}. Weyl gravity can represent an alternative to the standard dark matter paradigm, since it can explain the flat rotation curves of galaxies without introducing a dark matter component \cite{M1}.

 The importance of the conformal structures in cosmology was pointed out by Penrose \cite{P1}, who introduced a cosmological scenario called {\it Conformal Cyclic Cosmology} (CCC). The starting point of the Conformal Cyclic Cosmology is the observation that at the end of the de Sitter phase, the Universe will be space-like, and conformally flat. During the Big Bang, the initial boundary of the Universe did also have the same property. In the CCC model the Universe is formed of eons-time oriented spacetimes, with spacelike null infinities. The  Conformal Cyclic Cosmology model was investigated in \cite{P1a, P2,P3,P4,P5,P6,P7}.

The role of the local conformal symmetry was investigated by 't Hooft in \cite{H1}, based on the assumption that it is  {\it an exact symmetry of Nature, which is broken spontaneously}. The breaking of the conformal invariance may explain  the small-scale features of the gravitational force. It may also be possible that conformal symmetry is as important as the Lorentz invariance of the physical laws, and give important hints for the understanding of the physics of the Planck scale. Based on the idea that conformal invariance is locally an exact, but spontaneously broken symmetry of nature, a theory of gravity  was introduced in \cite{H2}.  In this approach, the conformal component of the metric field may be interpreted as a dilaton field. Moreover, black holes become regular solitons, topologically trivial, without singularities, horizons, and firewalls.

Weyl geometry have been extensively applied for the study of the gravitational phenomena in the framework of the $f(Q)$ gravity theory, or the symmetric teleparallel gravity. This theory was proposed in \cite{Q1}, under the assumption that the nonmetricity $Q$ of a Weyl geometry is the basic geometrical quantity that describes gravitational interaction. The initial formulation of \cite{Q1} was generalized in \cite{Q2}, thus leading to the $f(Q)$ gravity theory, or the nonmetric gravity theory. The action for the gravitational field is given by $S=\int{\left(f(Q)+L_m\right)\sqrt{-g}d^4x}$. The  cosmological, geometrical and physical properties of the $f(Q)$ theory have been investigated in \cite{Q2, Q3, Q4, Q5, Q6}.

Conformal Weyl geometric gravity, constructed ab initio in the framework of Weyl geometry, was studied, in both metric and Palatini formulations, in \cite{Gh1,Gh2,Gh3,Gh4,Gh5,Gh6,Gh7, Gh8,Gh9,Gh10, Gh11, Gh12, Gh13}. In this theory the gravitational action is quadratic in both the scalar Weyl curvature, and in the Weyl tensor. The implications of the theory for the evolution of the very early Universe, as well as for the elementary particle physics were investigated in detail. An important property of the quadratic Weyl geometric action is that it has spontaneous symmetry breaking in a Stueckelberg mechanism.  Consequently, the Weyl gauge field becomes massive. Hence, from the Weyl action one reobtains the Hilbert-Einstein action of general relativity, together with  a positive cosmological constant, and a Proca action type action for the massive Weyl gauge field \cite{Gh1}.

The coupling of matter and geometry in a conformally invariant way was studied in \cite{Gh10}, leading to the so-called conformally invariant $f\left(R,L_m\right)$ theory. The coupling term was assumed to have the form $L_m\tilde{R}^2$, where $L_m$ is the ordinary (baryonic) matter Lagrangian, and $\tilde{R}$ is the Weyl scalar. This type of coupling satisfies the requirement of conformal invariance. As is usual in theories with geometry-matter coupling,  the divergence of the matter energy-momentum tensor does not vanish, and an extra force, depending on the Weyl vector, and on the matter Lagrangian does appear in the equations of motion of the massive particles. The cosmological implications of the conformally invariant $f\left(R,L_m\right)$ theory were also considered for the case of the flat Friedmann-Lemaitre-Robertson-Walker geometry. The corresponding model can give a good description of the observational data for the Hubble function up to a redshift of the order of $z\approx 3$.

Numerical black hole solutions in the Weyl geometric gravity were considered in \cite{Gh13}, after reformulating the field equations in a dimensionless form, and after introducing an appropriate independent radial coordinate.  An exact black hole solution, representing the geometry in which the Weyl vector has only a radial spacelike component was also derived. By also using numerical methods, the thermodynamic properties of the Weyl geometric type black holes, including the horizon temperature, teh specific heat, the entropy, and the Hawking evaporation time were also analyzed in detail.

It is the goal of the present paper to investigate another physical aspect of  the Weyl geometric gravity theory, as introduced and developed in \cite{Gh1,Gh2,Gh3,Gh4,Gh5}, namely, the possibility that the behavior of the galactic rotation curves, usually explained through the presence of dark matter, could in fact be explained as a geometric effect in Weyl gravity. In order to analyze the behavior of the galactic rotation curves, and the possibility of their geometric explanation,  we introduce first a basic, conformally invariant gravitational model, whose action is constructed ab initio in Weyl geometry. We consider the simplest possible model, in which the action contains two terms only, the square of the Weyl scalar, and the strength of the Weyl vector, respectively. An equivalent scalar-vector-tensor theory can be obtained by introducing an auxiliary scalar field $\phi_0$, leading, finally, in the Riemann space to a theory containing a scalar field, nonminally coupled to the Ricci scalar, and to the Weyl vector field. Moreover, a Higgs type potential proportional to $\phi _0^4$ also does appear in the theory. We obtain the gravitational field equations, together with the generalized Klein-Gordon and Proca equations for this action in the metric formalism.

The static spherically symmetric solutions of the Weyl geometric gravity are investigated by assuming that the Weyl vector has only a nonzero radial component. For this case the gravitational field equations do admit an exact solution, which, interestingly enough, has a form very similar to the previous solution (\ref{solM}), obtained in the framework of Weyl gravity, despite the major differences between the two theories. We investigate the problem of the rotation curves in Weyl geometric gravity by obtaining the general expression of the tangential velocities in the theory. The presence of the scalar field in the gravitational field equations allows the definition of an equivalent geometric mass, and of a corresponding density profile, which can be interpreted as providing an effective description that could give a theoretical, purely geometrical explanation, of the   phenomenology usually explained through the presence of dark matter. Some specific models, corresponding to various particular forms of the Weyl vector, are also considered. Finally, in order to test the model, we perform a comparison of the theoretical with a set of observations of a small selected sample of rotation curves. In order to obtain realistic models the effects of the baryonic matter are also included. This comparison allows us to fix the constants that do appear in the description of the scalar field profile of the exact solution of the model.

The present paper is organized as follow. The action of the theory is written down in Section~\ref{sect1}, where the generalized Einstein, Klein-Gordon and Proca field equations are also obtained. The spherically symmetric vacuum field equations of the theory are presented in Section~\ref{sect2} for a particular choice of the Weyl vector. An exact analytic solution of the field equations is also obtained. The behavior of the tangential velocity of massive test particles in Weyl geometric gravity is analyzed in Section~\ref{sect3}, where, after obtaining the general expressions of the velocity, three simple particular cases, corresponding to some specific choices of the functional form of the Weyl vector, are investigated in detail. The role of the baryonic matter is considered in Section~\ref{conbrsec}, where a comparison between the theoretical predictions and a selected sample of observational galactic rotation curves is also performed. We discuss and conclude our results in Section~\ref{sect5}. The results of the fittings of a selected set of rotation curves is presented in Appendix~\ref{app}.

\section{Gravitational field equations in Weyl geometric conformal gravity}\label{sect1}

In the present Section we briefly introduce first the mathematical foundations of Weyl geometry, by pointing out its three main characteristics: variation of length of vectors during parallel transport, nonmetricity, and conformal invariance, respectively. Then we write down the action of the simplest gravitational theory, inspired by Weyl-geometry, and which is obtained by requiring the conformal invariance of the physical laws. This action is quadratic in the Weyl scalar, and it can be linearized with the help of an auxiliary scalar field. The gravitational action can be reformulated in the Riemann geometry, and by varying it with respect to the metric, we derive the gravitational field equations that give the evolution of the metric, of the scalar field, and of the Weyl vector, respectively.

\subsection{Basic concepts of Weyl geometry}

On manifold on which it is defined, the metric tensor allows to define distances and angles near each point. At a point $x$ of the manifold the metric tensor is a bilinear form defined on the tangent space at $x$. In the first non-Euclidian geometry, the Riemannian geometry \cite{Riemm}, under the parallel transport of a vector with respect to the connection, the length of a vector is preserved. The connection constructed in Riemann geometry, a metric geometry in which the covariant derivative of the metric identically vanishes, is called the Levi-Civita connection, or the Christoffel symbols $ \Gamma^\lambda_{\mu\nu}$, if we consider its tensorial components. However, a few years after Einstein and Hilbert formulated the correct field equations of general relativity, derived by Hilbert by using a variational principle, H. Weyl \cite{Weyl1, Weyl2} generalized Riemann geometry by assuming the existence of a conformal transformation in every point $x$ of the space-time manifold. For the  metric tensor $g_{\mu \nu}$, a conformal transformation $g_{\mu \nu}\rightarrow \tilde{g}_{\mu \nu}$ is defined  as
\be
\tilde{g}_{\mu \nu } =\Sigma ^{n}(x)g_{\mu \nu },
\ee
where $\Sigma (x)$ is an arbitrary function, and $n$ is called the Weyl charge. In the following we will limit our investigations to the case $n=1$. An essential feature of the Weyl geometry is that the length $l$ of a vector changes when it is parallely transported along an infinitesimal path,  from $x^\mu$ to $x^\mu + \delta
x^\mu$. The change in the length of a vector is given by
\begin{eqnarray}  \label{a1}
\delta l = l \omega_\mu \delta x^\mu,
\end{eqnarray}
where $\omega_\mu$ is called the Weyl vector field. The Weyl vector field is the source of the nonmetricity $Q_{\lambda \mu \nu}$ of the Weyl manifold, via the nonconservation of the covariant derivative of the metric tensor, which generally is given by
\begin{align}\label{nmetr}
& \tilde{\nabla}_\lambda g_{\mu\nu} = - \alpha \omega_\lambda g_{\mu\nu}=Q_{\lambda \mu \nu},
\end{align}
where $\alpha$ is called the Weyl gauge coupling constant. From the nonmetricity condition (\ref{nmetr}) one obtains in a straightforward way the connection of the Weyl geometry, which is given by
\begin{align}\label{connex}
& \tilde{\Gamma}^\lambda_{\mu\nu} = \Gamma^\lambda_{\mu\nu} + \frac{1}{2}
\alpha \Big[ \delta^\lambda_\mu \omega_\nu + \delta^\lambda_\nu \omega_\mu -
g_{\mu\nu} \omega^\lambda \Big].
\end{align}
Eq.~(\ref{connex}) represents a generalization of the standard Levi-Civita connection, associated to the metric $g$. In the following we denote by a tilde all geometrical and physical quantities defined in the Weyl geometry.

With the help of the Weyl connection we obtain the curvature tensor, defined as
\begin{equation}
\tilde{R}_{\mu \nu \sigma }^{\lambda }=\partial _{\nu }\tilde{\Gamma}_{\mu
\sigma }^{\lambda }-\partial _{\sigma }\tilde{\Gamma}_{\mu \nu }^{\lambda }+%
\tilde{\Gamma}_{\rho \nu }^{\lambda }\tilde{\Gamma}_{\mu \sigma }^{\rho }-%
\tilde{\Gamma}_{\rho \sigma }^{\lambda }\tilde{\Gamma}_{\mu \nu }^{\rho },
\end{equation}%
and its first contraction,
\begin{equation}
\tilde{R}_{\mu \nu }=\tilde{R}_{\mu \lambda \nu }^{\lambda },\tilde{R}%
=g^{\mu \sigma }\tilde{R}_{\mu \sigma },
\end{equation}%
respectively. The second contraction of the curvature tensor gives the Weyl scalar $\tilde R$,  which can be obtained as
\begin{eqnarray}
\tilde R = g^{\mu\nu} \Big( \partial_\rho \tilde{\Gamma}^\rho_{\mu\nu} -
\partial_\nu\tilde{\Gamma}^\rho_{\mu\rho} + \tilde{\Gamma}^\rho_{\mu\nu}
\tilde{\Gamma}^\sigma_{\rho\sigma} - \tilde{\Gamma}^\sigma_{\mu\rho}\tilde{%
\Gamma}^\rho_{\nu\sigma} \Big),
\end{eqnarray}
or, equivalently, as
\begin{equation}\label{R}
\tilde{R}=R-3n\alpha \nabla _{\mu }\omega ^{\mu }-\frac{3}{2}\left( n\alpha
\right) ^{2}\omega _{\mu }\omega ^{\mu },
\end{equation}
where $R$ denotes the Ricci scalar, defined in the Riemann geometry.

The Weyl tensor, which plays a central role in the conformally invariant gravitational theories \cite{M1,M2,M3,M4,M5,M6}, is defined according to
\begin{equation}
\tilde{C}_{\mu \nu \rho \sigma }^{2}=C_{\mu \nu \rho \sigma }^{2}+\frac{3}{2}%
\left( \alpha n\right) ^{2}\tilde{F}_{\mu \nu }^{2},
\end{equation}
where $C_{\mu \nu \rho \sigma }$ is the Weyl tensor, defined in the standard Riemannian geometry \cite{LaLi}.
$C_{\mu \nu \rho \sigma }^{2}$ is given, in terms of the Riemann and Ricci tensors, and of the Ricci scalar, by the relation
\begin{equation}
C_{\mu \nu \rho \sigma }^{2}=R_{\mu \nu \rho \sigma }R^{\mu \nu \rho \sigma
}-2R_{\mu \nu }R^{\mu \nu }+\frac{1}{3}R^{2}.
\end{equation}

After a conformal transformation with a conformal factor $\Sigma$ in a point $x$, the transformation laws of the
metric tensor, of the Weyl field, and of an arbitrary  scalar field $\phi $ are given by,
\begin{eqnarray}  \label{a2}
\hat g_{\mu\nu} = \Sigma g_{\mu\nu}, \hat \omega _\mu = \omega_\mu -
\frac{1}{\alpha} \partial_\mu \ln\Sigma, \hat \phi = \Sigma^{-\frac{1}{2%
}} \phi.
\end{eqnarray}

In the following, {\it we will adopt the view that the Weyl vector field is a purely geometric quantity, without any specific relation with the electromagnetic potential}.

Presently, most of the physical studies applying Weyl geometry are based on the basic hypothesis  {\it that conformal invariance represents a fundamental symmetry property of the physical laws}. A local conformal transformation
\begin{equation}
d\tilde{s}^2=\Sigma ^2(x)ds^2=\Sigma ^2(x)g_{\alpha \beta}dx^{\alpha}dx^{\beta}=\tilde{g}_{\alpha \beta}dx^{\alpha}dx^{\beta},
\end{equation}
 {\it does not lead to a transformation of the local coordinates}, but it {\it modifies the units of the measurements of the physical quantities} \cite{Weyl3}. Under a conformal transformation, the Christoffel symbols transform as \cite{Weyl3}
 \begin{equation}
 \tilde{\Gamma}^{\lambda}_{\alpha \beta}=\Gamma _{\alpha \beta}^{\lambda}+\Big[\frac{\Sigma_{,\alpha}}{\Sigma}\delta_{\beta}^{\lambda}+\frac{\Sigma _{,\beta}}{\Sigma}\delta_{\alpha}^{\lambda}-g^{\lambda \sigma}\frac{\Sigma_{,\sigma}}{\Sigma}g_{\alpha \beta}\Big].
 \end{equation}

 By taking into account the law of transformation of the Weyl vector,  $\tilde{\omega}_{\alpha}=\omega _{\alpha}+2\Sigma _{,\alpha}/\Sigma$, it follows that
 \be
 \tilde{\Gamma}^{\lambda}_{\alpha \beta}=\Gamma _{\alpha \beta}^{\lambda}, \tilde{R}^{\alpha}_{\beta \lambda \sigma}=R^{\alpha}_{\beta \lambda \sigma}, \tilde{R}_{\alpha \beta}=R_{\alpha \beta},
 \ee
 and
 \be
 \tilde{F}_{\alpha \beta}=F_{\alpha \beta},
 \ee
 respectively, where
 \be
 \tilde{F}_{\alpha\beta} =  \tilde{\nabla}_{[\alpha} \omega_{\beta]} =  \nabla_{[\alpha}
\omega_{\beta]} =  \partial_{[\alpha} \omega_{\beta]}=\partial _{\alpha}\omega _\beta-\partial _\beta \omega _\alpha,
\ee
is the Weyl field strength.

Therefore, one can interpret $\omega _{\mu}$ as {\it a gauge field mediating at different space-time points the conformal factors} \cite{Bera, Berb}.
\\
\subsection{Action and field equations}

The simplest conformally invariant Lagrangian density,  which can be constructed in Weyl geometry, is given by \cite{Weyl1, Weyl2, Gh3,Gh7}
\bea\label{inA}
L_0=\Big[\, \frac{1}{4!}\,\frac{1}{\xi^2}\,\tilde R^2  - \frac14\, F_{\mu\nu}^{\,2} \Big]\sqrt{-g},
\eea
where we assume that the perturbative coupling constant $\xi$ satisfies the condition  $\xi < 1$. The action (\ref{inA}) contains a scalar degree of freedom, which can be extracted by replacing in $L_0$ the term  $\tilde{R}^2$ by
\be
\tilde{R}^2\rightarrow -2 \phi_0^2\,\tilde R-\phi_0^4,
\ee
where $\phi_0$ represents {\it an (auxiliary)  scalar field}.

The Lagrangian density obtained by introducing the scalar field is equivalent with the initial one, since
after inserting the solution $\phi_0^2=-\tilde R$ of the equation of motion of $\phi_0$ in the new Lagrangian
$L_1$, we reobtain the initial action  Eq.~(\ref{inA}).

Therefore, by starting with the action (\ref{inA}), we obtain an equivalent Weyl geometric Lagrangian,
\bea\label{alt3}
L_1=\sqrt{-g} \Big[-\frac{1}{12}\frac{1}{\xi^2}\,\phi_0^2\,\tilde R
-\frac14 \,F_{\mu\nu}^2-\frac{\phi_0^4}{4!\,\xi^2}
\Big],
\eea
which is linear in the Weyl curvature $\tilde{R}$.

The gravitational Lagrangian (\ref{alt3}) represents the simplest possible physical Lagrangian density that fully contains  Weyl gauge symmetry, and conformal invariance, respectively.  An important property of $L_1$  is that it spontaneously
breaks down to an Einstein-Proca type Lagrangian of the Weyl gauge field.

By substituting in Eq.~(\ref{alt3})  $\tilde{R}$ by its Riemannian counterpart, as given by Eq.~(\ref{R}), after performing a gauge transformation that allows us the redefine the physical and geometrical variables, we obtain an action, defined in the Riemann space, invariant under conformal transformations,  and which is given by \cite{Gh3,Gh4,Gh5},
\begin{eqnarray}\label{a3}
\mathcal{S }&=& \int d^4x \sqrt{-g} \Bigg[ -\frac{1}{12} \frac{\phi^2}{\xi^2}
\Big( R - 3\alpha\nabla_\mu \omega^\mu - \frac{3}{2} \alpha^2 \omega_\mu
\omega^\mu \Big) \nonumber\\
&&- \frac{1}{4!}\frac{\phi^4}{\xi^2} - \frac{1}{4} F_{\mu\nu}
F^{\mu\nu} \Bigg].
\end{eqnarray}

We will call the theory described by the above action {\it the Weyl geometric gravity theory}, in order to differentiate it with respect to other various gravitational theories already proposed, and  mainly constructed from the Weyl tensor defined in Riemann geometry.   By varying the action Eq.~(\ref{a3}) with respect to the metric tensor we find the gravitational field
equations of the Weyl geometric gravitational theory \cite{Gh13},
\begin{eqnarray}\label{b2a}
&&\frac{\phi ^{2}}{\xi ^{2}}\Big(R_{\mu \nu }-\frac{1}{2}Rg_{\mu \nu }%
\Big)\nonumber\\
&&-\frac{3\alpha }{2\xi ^{2}}\Big(\omega ^{\rho }\nabla _{\rho }\phi
^{2}g_{\mu \nu }-\omega _{\nu }\nabla _{\mu }\phi ^{2}-\omega _{\mu }\nabla
_{\nu }\phi ^{2}\Big)\nonumber\\
&&+\frac{3\alpha ^{2}}{4\xi ^{2}}\phi ^{2}\Big(\omega
_{\rho }\omega ^{\rho }g_{\mu \nu }-2\omega _{\mu }\omega _{\nu }\Big)
+6F_{\rho \mu }F_{\sigma \nu }g^{\rho \sigma }-\frac{3}{2}F_{\rho \sigma
}^{2}g_{\mu \nu }\nonumber\\
&&-\frac{1}{4\xi ^{2}}\phi ^{4}g_{\mu \nu }
+\frac{1}{\xi ^{2}}%
\Big(g_{\mu \nu }\Box -\nabla _{\mu }\nabla _{\nu }\Big)\phi ^{2}=0.
\end{eqnarray}

By taking the trace of Eq.~(\ref{b2a}) we obtain,
\begin{equation}\label{b3n}
\Phi R+3\alpha \omega ^{\rho }\nabla _{\rho }\Phi +\Phi ^{2}-\frac{3}{2}%
\alpha ^{2}\Phi \omega _{\rho }\omega ^{\rho }-3\Box \Phi =0,
\end{equation}%
where we have denoted $\Phi \equiv \phi ^{2}$.

The variation of the action  Eq.~(\ref{a3})
with respect to the scalar field $\phi $ gives the equation of motion of $%
\Phi $,
\begin{equation}\label{b4}
R-3\alpha \nabla _{\rho }\omega ^{\rho }-\frac{3}{2}\alpha ^{2}\omega _{\rho
}\omega ^{\rho }+\Phi =0.
\end{equation}%

After multiplying Eq.~(\ref{b4}) by $\Phi$, and subtracting it  from Eq.~(\ref{b3n}), we find the generalized Klein-Gordon equation for the scalar field $\Phi$ as given by
\begin{equation}\label{b5}
\Box \Phi-\alpha \nabla _{\rho }(\Phi\omega ^{\rho })=0.
\end{equation}%

By varying the action Eq.~(\ref{a3}) with respect to $\omega _{\mu }$,
we obtain the equation of motion of the Weyl vector as,
\begin{equation}\label{Fmunu}
4\xi ^{2}\nabla _{\nu }F^{\mu \nu }-\alpha ^{2}\Phi\omega ^{\mu }+\alpha \nabla
^{\mu }\Phi=0.
\end{equation}%

By applying to both sides of Eq.~(\ref{Fmunu}) the operator  $\nabla _{\mu }$, we obtain again Eq.~(\ref{b5}), a result that indicates the consistency of the Weyl geometric gravity  theory.

\section{The vacuum field equations in spherical symmetry}\label{sect2}

 In the following our main goal is to investigate the motion of massive particles in Weyl geometric gravity in {\it a static and spherically symmetric geometry}, in which all quantities depend only on the radial coordinate $r$. In the following we adopt a system of coordinates given by $\left(t,r,\theta, \varphi\right)$, where $t$ is the time, $r$ denotes the radial coordinate, while $\theta$ and $\varphi$ are the angular coordinates. We also adopt the natural system of units, with $c=G=1$.
Therefore, in the static spherically symmetric geometry the space-time interval can be written down as,
\begin{equation}\label{line}
ds^{2}=e^{\nu (r)}dt^{2}-e^{\lambda (r)}dr^{2}-r^{2}d\Omega ^{2},
\end{equation}%
where $d\Omega ^{2}=d\theta ^{2}+\sin ^{2}\theta d\varphi ^{2}$. Hence,  the
components of the metric tensor are given by,
\begin{equation}
g_{\mu \nu }=\mathrm{diag}(e^{\nu (r)},-e^{\lambda (r)},-r^{2},-r^{2}\sin
^{2}\theta ).
\end{equation}

In a spherically symmetric gravitational field in the presence of Weyl geometric effects,  the third and the
fourth components of the Weyl vector  must vanish, due to the choice of the metric. Therefore, generally, in spherical symmetry, the Weyl
vector has the components $\omega _{\mu }=\left( \omega _{0},\omega
_{1},0,0\right) $ \cite{Gh13}.

\subsection{Static spherically symmetric field equations for $\omega _{\mu }=\left( 0,\omega %
_{1},0,0\right) $}

In the present study we restrict our analysis of the Weyl geometric gravity to the case in which the temporal component, $\omega _{0}$,
of the Weyl vector vanishes identically. Therefore, the Weyl vector $\omega _{\mu}$ takes the form
\be
\omega _{\mu }=(0,\omega _{1},0,0),
\ee
containing a {\it non-zero radial component only}. For this simple form of $\omega _{\mu}$, the strength of the Weyl vector identically vanishes,   $F_{\mu \nu}\equiv 0$. Therefore, from Eq.~(\ref{Fmunu}) we immediately obtain
\be\label{15}
\Phi '=\alpha \Phi \omega _1.
\ee

Since
\be
\Box \Phi=%
\frac{1}{\sqrt{-g}}\frac{\partial }{\partial x^{\alpha }}\left( \sqrt{-g}%
g^{\alpha \beta }\frac{\partial \Phi}{\partial x^{\beta }}\right) ,
\ee
and
\be
\nabla _{\alpha }\omega ^{\alpha }=\frac{1}{\sqrt{-g}}\frac{\partial }{\partial
x^{\alpha }}\left( \sqrt{-g}\omega ^{\alpha }\right),
\ee
respectively, Eq.~(\ref{b5}) becomes
\bea\label{16}
&&\frac{1}{\sqrt{-g}}\frac{d}{dr}\left( \sqrt{-g}g^{11}\frac{d\Phi}{dr}\right)
-\alpha \omega ^{1}\frac{d\Phi}{dr}\nonumber\\
&&-\alpha \Phi\frac{1}{\sqrt{-g}}\frac{d}{dr}%
\left( \sqrt{-g}\omega ^{1}\right) =0.
\eea

To make Eqs.~(\ref{15}) and (\ref{16}) mathematically consistent to each other, we consider the gauge condition on the Weyl vector, which
can be written down in the form,
\begin{equation}
\nabla _{\alpha }\omega ^{\alpha }=\frac{1}{\sqrt{-g}}\frac{\partial }{%
\partial x^{\alpha }}\left( \sqrt{-g}\omega ^{\alpha }\right) =0.
\end{equation}%

Hence,  in static spherically symmetric Weyl geometric gravity, as well as for the present choice of $\omega _{\mu}$, we immediately obtain the important relation
\begin{equation}
\frac{1}{\sqrt{-g}}\frac{d}{dr}\left( \sqrt{-g}\frac{d\omega ^{1}}{dr}%
\right) =0,
\end{equation}
leading to,
\bea\label{omeg1}
\hspace{-0.6cm}\omega ^{1}=\frac{C_{1}}{\sqrt{-\hat{g}}}=C_{1}\frac{e^{-{(\nu +\lambda )}/2}%
}{r^{2}},
\omega _{1}=-C_{1}\frac{e^{-\left( \nu -\lambda
\right) /2}}{r^{2}},
\eea%
where $C_{1}$ is an arbitrary constant of integration. Moreover, to simplify the mathematical formalism  we have denoted $%
\sqrt{-\hat{g}}=e^{\left( \nu +\lambda \right) /2}r^{2}$. Next we apply 
the gauge condition, and we use the specific form of $\omega ^{1}$. Then Eq.~(%
\ref{16}) gives \cite{Gh13},
\be\label{eqev}
\frac{1}{\sqrt{-g}}\frac{d}{dr}\left( \sqrt{-g}g^{11}\frac{d\Phi}{dr}\right)
-\alpha \omega ^{1}\frac{d\Phi}{dr}=0.
\ee

Alternatively, we have  
\begin{equation}\label{31}
\frac{1}{\sqrt{-\hat{g}}}\frac{d}{dr}\left( \sqrt{-\hat{g}}g^{11}\frac{d\Phi}{dr%
}\right) -\alpha \frac{C_{1}}{\sqrt{-\hat{g}}}\frac{d\Phi}{dr}=0.
\end{equation}%

From Eq.~(\ref{31}) we immediately obtain,
\begin{equation}\label{32}
\sqrt{-\hat{g}}g^{11}\frac{d\Phi}{dr}-\alpha C_{1}\Phi=C,
\end{equation}%
where $C$ is an arbitrary integration constant. To make the above equation consistent  with Eq.~(\ref{15}), $C$ must be taken as zero. Thus,  from Eq.~(\ref{32}) we find again Eq.~(\ref{15}), given by
$\Phi ^{\prime }=\alpha \omega _{1}\Phi$.

With the use of the explicit expression of $\omega ^{1}
$, as given by Eq.~(\ref{omeg1}), for the unknown function $\Phi$ we arrive at the first order ordinary differential equation,
\begin{equation}
\frac{d\Phi}{dr}=-C_{1}\frac{e^{-\left( \nu -\lambda \right) /2}}{r^{2}}\Phi.
\end{equation}


The gravitational field equations of Weyl geometric gravity (\ref{b2a}) give the following basic relations \cite{Gh13},
\bea\label{c15}
&& -1+e^{\lambda }-\frac{1}{4}e^{\lambda }r^{2}\Phi-\frac{2r\Phi^{\prime }}{\Phi}+%
\frac{3r^{2}}{4}\frac{\Phi^{\prime 2}}{\Phi^{2}}+r\lambda ^{\prime }\left(
1+\frac{r}{2}\frac{\Phi^{\prime }}{\Phi }\right)\nonumber\\
&&-\frac{r^{2}\Phi^{\prime \prime
}}{\Phi}=0,
\eea
\bea \label{c16}
&& 1-e^{\lambda }+\frac{1}{4}e^{\lambda }r^{2}\Phi+\frac{2r\Phi^{\prime }}{\Phi}+\frac{%
3r^{2}}{4}\frac{\Phi^{\prime 2}}{\Phi^{2}}+r\nu ^{\prime }\left(1+\frac{r}{2}\frac{\Phi'}{\Phi}\right)
=0,\nonumber\\
\end{eqnarray}
and, finally, 
\bea\label{e22}
\hspace{-0.5cm}&&2(\nu ^{\prime
}-\lambda ^{\prime })+(4-2r\lambda ^{\prime }+2r\nu ^{\prime })\frac{\Phi ^{\prime }}{\Phi }\nonumber\\
\hspace{-0.5cm}&&+r\left(e^{\lambda }\Phi +4\frac{\Phi ^{\prime \prime }}{\Phi }-3\frac{\Phi
^{\prime 2}}{\Phi ^{2}}-\lambda ^{\prime }\nu ^{\prime }+\nu ^{\prime
2}+2\nu ^{\prime \prime }\right)=0.
\eea

It can be immediately shown that Eq.~\eqref{e22} is a consequence of Eqs.~(\ref{c15}) and (\ref{c16}), obtained through their linear combination. From the sum of  Eqs.~(\ref{c15}) and (\ref{c16}), we find an evolution equation for the auxiliary scalar field of Weyl geometric gravity in spherical symmetry, given by \cite{Gh13},
\be\label{sum}
\frac{\Phi ''}{\Phi}-\frac{3}{2}\frac{\Phi ^{\prime 2}}{\Phi ^2}-\frac{\nu '+\lambda '}{r}\left(1+\frac{r}{2}\frac{\Phi '}{\Phi}\right)=0.
\ee

\subsection{Exact solution of the field equations with radial component of the Weyl vector field}

The system of the field equations (\ref{c15})-(\ref{sum}) of the static spherically symmetric Weyl geometric gravity does admit an exact solution, which can be obtained as  follows. In order to obtain the exact solution we assume first that the two independent metric tensor components $\nu $ and $\lambda$ satisfy the condition,
\be
\nu (r)+\lambda (r)=0,
\ee
a condition which we assume is valid for all $r>0$.

Then, from Eq.~(\ref{sum}), we straightforwardly find for $\Phi$ the second order ordinary differential equation,
\be
\Phi''=\frac{3}{2}\frac{\Phi ^{\prime 2}}{\Phi},
\ee
having the general solution represented by the function \cite{Gh13},
\be\label{40}
\Phi(r)=\frac{C_1}{\left(r+C_2\right)^2},
\ee
where $C_1$ and $C_2$ are arbitrary integration constants, to be determined from the initial or the boundary conditions..
The scalar field has the property
\be
\lim_{r\rightarrow \infty}\Phi(r)=0,
\ee
which indicates that $\Phi$ vanishes at infinity.
 Eq.~(\ref{c15}) can be rewritten as
\bea\label{41}
&&-1+\frac{d}{dr}\left(re^{-\lambda}\right)+\frac{1}{4}r^2\Phi+2re^{-\lambda}\frac{\Phi'}{\Phi}-\frac{3}{4}r^2e^{-\lambda}\frac{\Phi ^{\prime 2}}{\Phi ^2}\nonumber\\
&&-\frac{r^2\lambda 'e^{-\lambda}}{2}\frac{\Phi'}{\Phi}+r^2e^{-\lambda}\frac{\Phi ''}{\Phi}=0.
\eea

Now by making use of the mathematical identity,
\be
-\frac{r^2\lambda 'e^{-\lambda}}{2}\frac{\Phi'}{\Phi}=-\frac{re^{-\lambda}}{2}\frac{\Phi'}{\Phi}+\frac{r}{2}\frac{d}{dr}\left(re^{-\lambda}\right)\frac{\Phi'}{\Phi},
\ee
Eq.~(\ref{41}) can be reformulated as \cite{Gh13},
\bea\label{58a}
&&\left(1+\frac{r}{2}\frac{\Phi '}{\Phi}\right)\frac{d}{dr}\left(re^{-\lambda}\right)+ \left(\frac{3}{2}\frac{\Phi'}{\Phi}-\frac{3}{4}r\frac{\Phi^{\prime 2}}{\Phi ^2}+r\frac{\Phi ''}{\Phi}\right)\left(re^{-\lambda}\right)\nonumber\\
&&+\frac{1}{4}r^2\Phi-1= 0.
\eea

By taking into account the expression of $\Phi (r)$ as given by Eq.~(\ref{40}), we immediately find the differential equation,
\bea\label{44}
&&\left(1-\frac{r}{C_2+r}\right)
   u'(r)+3\left[\frac{r}{(C_2+r)^2}-\frac{1}{C_2+r}\right] u(r)\nonumber\\
 &&  +\frac{C_1 r^2}{4 (C_2+r)^2}-1=0,
\eea
where we have introduced the notation $u=re^{-\lambda}$. Eq.~(\ref{44}) is a first order linear differential equation, and its solution can be obtained through standard integration procedures. Therefore, we immediately find for $e^{-\lambda}$ the expression \cite{Gh13},
\bea\label{58}
re^{-\lambda}&=&\frac{r^2 \left(12 C_3 C_2^2+C_1-4\right)}{4 C_2}+r \left(3 C_3
   C_2^2+\frac{C_1}{4}-2\right)\nonumber\\
  && +C_3 C_2^3+\frac{1}{12}
   (C_1-12) C_2+C_3 r^3,
\eea
where by $C_3$ we have denoted an arbitrary integration constant.

It is interesting to note that various relations between the integration constants $C_1$, $C_2$ and $C_3$, lead to the possibility of representing the  metric (58) in two different forms. If one assumes that the integration constants satisfy the relation
 $C_3C_2^3+C_1C_2/12=C_2$, then the metric (\ref{58}) takes the simple form,
\be
e^{-\lambda}=e^{\nu}=1+\frac{2}{C_2}r+C_3r^2.
\ee
This metric gives an extension of the static cosmological de Sitter solution, with the Schwarzschild term proportional to $1/r$ missing from the line element. Moreover, the metric is not asymptotically flat. On the other hand, if assumes that the integration constants satisfy the relation, 
\be
3 C_3C_2^2+\frac{C_1}{4}=2(1+\beta),
\ee
where by $\beta $ we have denoted  an arbitrary constant, the metric tensor components of the line element describing the static spherically symmetric vacuum in Weyl geometric gravity take the form \cite{Gh13},
\be
e^{-\lambda}=e^{\nu}=2\beta +\frac{1+2\beta}{C_2}r-\frac{C_2\left(1-2\beta\right)}{3}\frac{1}{r}+C_3r^2.
\ee

If, by exploiting the analogy with the Schwarzschild metric,  we assume that $C_2\left(1-2\beta\right)/3=r_g$, where by $r_g$ we have denoted the gravitational radius of the massive object, the metric tensor components (\ref{58}) become \cite{Gh13},
\be\label{dm1}
e^{-\lambda}=e^\nu=2\beta +\frac{1-4\beta ^2}{3}\frac{r}{r_g}-\frac{r_g}{r}+C_3r^2.
\ee

Finally, by denoting $1-2\beta=\delta$, the expression of the metric (\ref{dm1}) can be reformulated as,
\be\label{dmef}
e^{-\lambda}=e^\nu=1-\delta +\frac{\delta(2-\delta)}{3}\frac{r}{r_g}-\frac{r_g}{r}+C_3r^2.
\ee

The behavior of the radial component of the Weyl vector is obtained as
\be
\omega _1=\frac{1}{\alpha}\frac{\Phi '}{\Phi}=-\frac{2}{\alpha}\frac{1}{r+C_2},
\ee
and has the property $\lim_{r\rightarrow \infty}\omega _1(r)=0$.

\section{Rotational velocity of massive particles in geometric Weyl gravity}\label{sect3}

We investigate now the problem of describing the physical properties of the stable circular
timelike geodesic orbits of massive test particles in the geometric Weyl gravity, by assuming that the gravitational field is static, and spherically
symmetric, having the metric given in a general form by Eq. (\ref{line}).

\subsection{Rotational velocity in static spherical symmetry}

In an arbitrary static and spherically symmetric metric the motion of a massive test particle is described by the Lagrangian \cite{Ma03a,Ma03b,Ma03c}
\begin{equation}
2L_{p}=e^{\nu \left( r\right)
}\left(
\frac{dt}{ds }\right) ^{2}-e^{\lambda \left( r\right) }\left( \frac{dr}{%
ds }\right) ^{2}-r^{2}\left( \frac{d\Omega }{ds }\right)
^{2}, \label{lag}
\end{equation}
where we have chosen the interval $s$ as the affine parameter along the
geodesics. In the timelike case, $s$ also corresponds to the proper
time. In the following,  by an overdot we denote the differentiation
with respect to $s $.

From the particle Lagrangian given by Eq.~(\ref{lag}) we immediately obtain two important conserved quantities, the energy $E=e^{\nu }\dot{t}={\rm constant}$, and the $%
\varphi $-component $l_{\varphi }=r^{2}\sin ^{2}\theta
\dot{\varphi}={\rm constant}$ of the
angular momentum of the particle. However, the $\theta $-component of the angular momentum, $%
l_{\theta }=r^{2}\dot{\theta}$ is not conserved,
but the total angular momentum $l^{2}=l_{\theta }^{2}+\left(
l_{\varphi }/\sin \theta \right) ^{2}$ is a constant of the motion,
$l^{2}={\rm constant}$.
The total angular momentum can be obtained as $l^{2}=r^{4}%
\dot{\Omega}^{2}$ \cite{Ma03a,Ma03b,Ma03c}.

The equation of the geodesic orbits (the equation of motion of the particle), can be
written in the timelike case as
\begin{equation}
\dot{r}^{2}+V\left( r\right) =0,
\end{equation}
where the potential $V\left( r\right) $ can be obtained as
\begin{equation}
V\left( r\right) =-e^{-\lambda }\left( E^{2}e^{-\nu }-\frac{l^{2}}{r^{2}}%
-1\right) .
\end{equation}

The case of the radial motion of particles in  stable circular orbits corresponds
to the conditions $\dot{r}=0$ and $\partial V/\partial r=0$, respectively,
with the motion taking place in an extremum of the potential $V\left( r\right)$. For the extremum to be a minimum we need also to impose the condition $\partial
^{2}V/\partial r^{2}>0$. If these three conditions are satisfied, then it follows  that the circular motion of test particles is stable. They also allow to determine the energy and the total angular momentum of
the particles as  \cite{LaLi, Ma03a,Ma03b,Ma03c},
\begin{equation}
E^{2}=\frac{2e^{\nu }}{2-r\nu ^{\prime }},l^{2}=\frac{r^{3}\nu ^{\prime }}{%
2-r\nu ^{\prime }}.  \label{cons}
\end{equation}

The metric given by Eq. (\ref{line}),
can be reformulated in terms of the spatial components of the
velocity as $ds^{2}=dt^{2}\left( 1-v^{2}\right) $, where we have denoted
\begin{equation}
v^{2}=e^{-\nu }\left[ e^{\lambda }\left( \frac{dr}{dt}\right)
^{2}+r^{2}\left( \frac{d\Omega }{dt}\right) ^{2}\right] .
\end{equation}

If the orbit is circular, $r={\rm constant}$, $\dot{r}=0$, and the tangential
velocity of the massive test particle is obtained according to
\begin{equation}
v_{tg}^{2}=\frac{r^{2}}{e^{\nu }}\left( \frac{d\Omega }{dt}\right)
^{2}.
\end{equation}

With the use of the conserved quantities, the angular velocity can be expressed as
by
\begin{equation}\label{vtgeq}
v_{tg}^{2}=\frac{e^{\nu }}{r^{2}}\frac{l^{2}}{E^{2}},
\end{equation}
or, equivalently,
\be
v_{tg}^{2}=\frac{r\nu ^{\prime }}{2}.
\ee
Thus, we have arrived to the basic result that the rotational velocity of the massive test particles  is completely determined by
the $g_{00}$ metric tensor component, $\exp \left( \nu \right) $, only.

\subsection{Tangential velocity in Weyl geometric gravity}

With the use of Eq.~(\ref{c16}) we immediately obtain for the tangential velocity of massive particles in Weyl geometric gravity, with the effects of the baryonic matter neglected, the general expression
\begin{equation}
v_{tg}^{2}=\frac{e^{\lambda }-1+e^{\lambda }r^{2}\Phi ^{2}/4-r\left[ 2\Phi
^{\prime }/\Phi +\left( 3r/4\right) \Phi ^{\prime 2}/\Phi ^{2}\right] }{2%
\left[ 1+\left( r/2\right) \left( \Phi ^{\prime }/\Phi \right) \right] }.
\end{equation}

By representing $e^{-\lambda }$ as
\be
e^{-\lambda }=1-\frac{2M_{DM}(r)}{r},
\ee
where $M_{DM}(r)$ can be interpreted as an effective geometric mass originating
from Weyl geometric effects, the tangential velocity becomes
\bea
v_{tg}^{2}&=&\frac{2M_{DM}(r)/r+r^{2}\Phi ^{2}/4}{2\left[ 1-2M_{DM}(r)/r\right]
\left[ 1+\left( r/2\right) \left( \Phi ^{\prime }/\Phi \right) \right] }\nonumber\\
&&-\frac{2\Phi ^{\prime }/\Phi +\left( 3r/4\right) \Phi ^{\prime 2}/\Phi ^{2}}{2%
\left[ 1+\left( r/2\right) \left( \Phi ^{\prime }/\Phi \right) \right] }.
\eea

The metric function can be obtained generally from Eq. (\ref{58a}), whose
general solution is given by
\bea
\hspace{-0.5cm}re^{-\lambda }&=&e^{\int f\left( \Phi (r),r\right) dr}\times \nonumber\\
\hspace{-0.5cm}&&\left\{ C+\int \frac{%
\left[ 1-r\Phi ^{2}/4\right] e^{-\int f\left( \Phi (r),r\right) dr}dr}{%
1+\left( r/2\right) \left( \Phi ^{\prime }/\Phi \right) }\right\} ,
\eea
where $C$ is an arbitrary integration constant, and we have denoted
\begin{equation}
f\left( \Phi (r),r\right) =\left( -\frac{3}{2}\frac{\Phi ^{\prime }}{\Phi }+%
\frac{3r}{4}\frac{\Phi ^{\prime 2}}{\Phi ^{2}}-r\frac{\Phi ^{\prime \prime }%
}{\Phi }\right) .
\end{equation}

Hence, for the effective geometric dark matter mass we obtain the general
expression
\bea
M_{DM}(r)&=&\frac{r}{2}\Bigg\{ 1-\frac{C}{r}e^{\int f\left( \Phi (r),r\right)
dr}-\frac{1}{r}e^{\int f\left( \Phi (r),r\right) dr}\times \nonumber\\
&&\int \frac{\left[
1-r\Phi ^{2}/4\right] e^{-\int f\left( \Phi (r),r\right) dr}dr}{1+\left(
r/2\right) \left( \Phi ^{\prime }/\Phi \right) }\Bigg\} .
\eea

We can also formally associate a geometric density profile to the ``dark matter" distribution, defined as
\be
\rho _{DM}(r)=\frac{1}{4\pi r^2}\frac{dM_{DM}(r)}{dr}.
\ee

\subsubsection{The case of the constant Weyl vector}

As a simple application of the Weyl geometric gravitational  theory we consider the case in which the radial component of the Weyl vector can be taken as a constant, $\omega_1\approx {\rm constant}<0$, at least in a finite region outside the baryonic distribution of the galaxy. Once the Weyl vector is fixed, the scalar field can be immediately obtained from Eq.~(\ref{15}) as
\be
\Phi (r)=\Phi_0e^{-\alpha \omega_1r},
\ee
where $\Phi_0$ is an arbitrary constant. Then Eq.~(\ref{58a}) takes the form
\bea
&&2\left(2- \alpha   \omega _1r\right) u'(r)+\alpha  \omega _1 \left(\alpha  \omega _1r-6\right) u(r)\nonumber\\
&&+r^2 \Phi_0 e^{-\alpha   \omega _1r}-4=0,
\eea
with the general solution given by
\be
re^{-\lambda}=\frac{e^{-\alpha  \omega _1r} \left(3 c_1 e^{\frac{3 \alpha  r \omega _1}{2}}-4  r^3 \Phi _0+48 r e^{\alpha   \omega _1r}\right)}{12\left (\alpha
   \omega _1r-2\right)^2},
\ee
where $c_1$ is an arbitrary constant of integration. We also assume that the range of the validity of the solution is for $r>2/\left(\alpha \omega_1\right)$.  The effective geometric mass of the dark matter can then be obtained as
\bea
\hspace{-0.5cm}&&M_{DM}(r)=\nonumber\\
\hspace{-0.5cm}&&\frac{r}{2} \left[1-\frac{e^{-\alpha  \omega _1r} \left(3 c_1 e^{\frac{3 \alpha   \omega _1r}{2}}-4  r^3 \Phi _0+48 r e^{\alpha  \omega _1r}\right)}{12 r \left(\alpha
   \omega _1r-2\right)^2}\right].
\eea

For the case of the constant Weyl vector, we obtain the effective dark matter density in the form
\begin{eqnarray}
\rho _{DM}(r)&=&-\frac{\alpha  c_1 \omega _1 e^{\frac{\alpha   \omega _1r}{2}} \left(\alpha   \omega _1r-6\right)}{64 \pi  \left(\alpha  r \omega _1r^2-2\right)^3}\nonumber\\
&&-\frac{\Phi _0 e^{-\alpha
    \omega _1r} \left(\alpha ^2  \omega _1^2r^2-3 \alpha   \omega _1r+6\right)}{24 \pi  \left(\alpha   \omega _1r-2\right)^3}\nonumber\\
 &&  +\frac{\alpha  \omega _1 \left(\alpha ^2
   \omega _1^2r^2-6 \alpha   \omega _1r+16\right)}{8 \pi   \left(\alpha   \omega _1r-2\right)^3}.
\end{eqnarray}

In order for the condition $\lim_{r\rightarrow \infty}\rho _{DM}(r)=0$ to be satisfied, we need to take the arbitrary integration constant $c_1$ as zero, $c_1=0$.  Finally, for the tangential velocity of massive test particles orbiting around the galactic center in a Weyl geometry in the presence of the radial component of the Weyl vector only we obtain the expression
\begin{equation}\label{86}
v_{tg}^{2}(r)=\frac{\alpha \omega _{1}}{2}r+\frac{\Phi _{0}r^{2}\left( \alpha
\omega _{1}r-2\right) }{2\left( 12e^{\alpha \omega _{1}r}-r^{2}\Phi
_{0}\right) }.
\end{equation}

The tangential velocity in this model increases with increasing $r$, and there is no limiting value of the speed of the particles.

\subsubsection{The case $\omega _1=1/(\alpha r)$}

We consider now a Weyl geometric type model in which the radial component of the Weyl vector is given by
\be
\omega _1 (r)=\frac{1}{\alpha r},
\ee
corresponding to a radially linearly increasing scalar field, given by
\be
\Phi(r)=\Phi_0r,
\ee
where $\Phi_0$ is an arbitrary constant. Then, for this choice of the Weyl vector, Eq.~(\ref{58a}) becomes
\be
u'(r)-\frac{1}{2r}u(r)+\frac{r}{2\alpha } -2=0,
\ee
with the general solution given by
\be
re^{-\lambda}=c_1 \sqrt{r}-\frac{r^2}{3 \alpha }+4 r,
\ee
where $c_1$ is an arbitrary constant of integration. The effective geometric dark matter mass is obtained as
\be\label{91}
M_{DM}(r)=\frac{r^2-9 \alpha  r-3 \alpha  c_1 \sqrt{r}}{6 \alpha },
\ee
with the corresponding effective dark matter density profile given by
\be
\rho _{DM}(r)=\frac{1}{8\pi}\left(\frac{2 r-9 \alpha }{3  \alpha  r^2}-\frac{c_1}{2  r^{5/2}}\right).
\ee
Finally, we obtain the tangential velocity of massive test particles in stable circular orbits as
\be
v_{tg}^2=\frac{1}{4}\left[1-\frac{3 \sqrt{r} (r-4 \alpha )}{ \left(3 \alpha  c_1-r^{3/2}+12 \alpha  \sqrt{r}\right)}\right].
\ee
In the limit of large distances, in this model the tangential velocity of test particles in Weyl geometric gravity tends towards the speed of light.

\subsubsection{Tangential velocity for the exact vacuum solution in Weyl geometric gravity}

For the exact solution of the Weyl geometric gravity theory, given by Eq.~(\ref{dmef}), the tangential velocity of test particles can be obtained immediately from its definition, and it is given by
\be
v_{tg}^2(r)=\frac{r \left[2 C_3 r+\frac{r_g}{r^2}+\frac{(2-\delta ) \delta }{3
   r_g}\right]}{2 \left[C_3 r^2-\delta +\frac{(2-\delta ) \delta  }{3
   r_g}r-\frac{r_g}{r}+1\right]}.
\ee

Generally, in the limit $r\rightarrow \infty$, $v_{tg}^2$ tends to 1, $\lim_{r\rightarrow \infty}v_{tg}^2(r)=1$. If one neglects the quadratic term in the metric, containing the integration constant $C_3$,  then $\lim_{r\rightarrow \infty}v_{tg}^2(r)=1/2$. The effective geometric dark matter profile can be obtained as,
\be
M_{DM}(r)=\frac{1}{2} r \left[-C_3 r^2+\delta -\frac{(2-\delta ) \delta  }{3
   r_g}r+\frac{r_g}{r}\right],
\ee
with the associated geometric dark matter density profile obtained as
\be
\rho_{DM}(r)=\frac{1}{24 \pi  r^2}\left(3 \delta  +2 (\delta -2) \delta  \frac{r}{r_g} -9 C_3 r^2\right).
\ee

In the limit $r\rightarrow \infty$ we obtain $\lim_{r\rightarrow \infty} \rho_{DM}(r)=-3C_3/8\pi$, which imposes the condition $C_3<0$ on the integration constant $C_3$.

\section{Galactic rotation curves in Weyl geometric gravity-comparison with observations}\label{conbrsec}

In the present Section we consider the behavior of the rotation curves in the Weyl geometric gravity model, introduced in the previous Sections. In order to compare the theoretical predictions with the observations we need to add a matter component to the model. Adding the matter component breaks the conformal invariance of the theory, but this effect is only important in the baryonic matter dominated regions of the galaxy. Once the baryonic model is added, one can proceed to fit the theoretical model with the observations.

\subsection{Conformal symmetry breaking by matter coupling}

In order to find the General Relativistic limit of the Weyl geometric gravity theory in the presence of a radial component of the Weyl vector, we insert a matter source into the gravitational action. This matter source will explicitly break the conformal invariance. We assume that the breaking of the conformal invariance can be described by the addition into the total gravitational action (\ref{a3}) of a matter-scalar field coupling term of the form
\be
\mathcal{ S}_{M}=-\int~d^{4}x\sqrt{-g}\frac{\phi^{2}}{12\xi^{2}}\kappa\mathcal{L}_{M},  \label{conbrL}
\ee
where $\mathcal{L}_{M}$ is the Lagrangian density of the ordinary matter.  By assuming that the baryonic matter can be represented as a perfect fluid, characterized by two thermodynamic quantities only, the thermodynamic pressure $P$, and the density $\rho$, then  the Lagrangian density has the following form \cite{Minazzoli:2012md},
\be\label{74}
\mathcal{L}_{M}=-\rho \left( 1+\int \frac{P(\rho)}{\rho^{2}}~d\rho \right),
\ee
where we have also assumed that the cosmic fluid satisfies a barotropic equation of state, $P=P(\rho)$. From the Lagrangian density (\ref{74}) we can obtain the energy-momentum tensor of the ideal baryonic fluid that sources the right-hand side~(RHS) of the field equation (\ref{b2a}) with a term of the form $\kappa T_{\mu\nu}\phi^{2}/\xi^{2}$.  This will result in the presence of a matter source term
\be
\kappa\Phi T = \kappa\Phi (\rho-3P),
\ee
in the RHS of Eq.~(\ref{b3n}), and of a term $-\kappa\mathcal{L}_{M}$ in the RHS of Eq.~(\ref{b4}).  The RHS of Eq.~(\ref{b5}) in the presence of matter thus becomes $\kappa\Phi (3P + \rho\int P/\rho^{2}~d\rho)$.

Remarkably, for fluids with negligible pressure, such as the baryonic matter in the galaxies, the scalar field profile will be almost identical to the vacuum profile, with no sources. In this case, the only modification induced by the presence of matter source is in Eq.~(\ref{44}), while the scalar profile is unchanged. Hence, by adding the baryonic matter term source to Eq.~(\ref{44}), we obtain
\bea\label{44s}
&&\frac{C_{2}}{C_2+r}
   u'(r)-\frac{3 C_{2}}{(C_2+r)^2} u(r)+\frac{C_1 r^2}{4 (C_2+r)^2}-1= -\kappa\rho r^{2}.  \notag \\
\eea

The general solution of Eq.~(\ref{44s}) is given by
\bea\label{sols}
re^{-\lambda}&=&\frac{r^2 \left(12 C_3 C_2^2+C_1-4\right)}{4 C_2}+r \left(3 C_3
   C_2^2+\frac{C_1}{4}-2\right)\nonumber\\
  && +C_3 C_2^3+\frac{1}{12}
   (C_1-12) C_2+C_3 r^3  \notag  \\
  && -\kappa \Big( 1+\frac{r}{C_{2}} \Big)^{3}\int_{0}^{r}\frac{\rho r^{2}}{\left( 1+\displaystyle{\frac{r}{C_{2}}}\right)^{2}}~dr.
\eea

The General Relativistic limit is given by the limits of small $r$, and large $C_{2}$, or, equivalently, for  $r/C_{2}\ll 1$. In this limit, the last term on the RHS of Eq.~(\ref{sols}) becomes $-\kappa \int \rho r^{2}~dr = -\kappa M(r)/4\pi$, and the solution of Eq.~(\ref{sols}) can be expressed as
\be\label{meteq}
e^{-\lambda}=e^\nu=1-\delta -\frac{1}{r}\Big[r_{g}+\frac{\delta(2-\delta)}{3\gamma}\Big]+\gamma r -\frac{\Lambda}{3} r^2,
\ee
where we have denoted
\be
\delta =3-3 C_{3}C_{2}^{2}-\frac{C_{1}}{4},~\gamma = \frac{2-\delta}{C_{2}},~\Lambda = -3C_{3},
\ee
where $r_{g}=2GM(r)/c^{2}$~(for $\kappa = 8\pi G/c^{4}$) is the Schwarzschild radius of the mass extending up to the radius $r$. Since the constraints on the Newtonian and General Relativistic limits for the Schwarzschild term $\sim 1/r$ are significantly stricter, the simplifying assumption $\delta = 0$ can be imposed without any loss of generality. In this case,
\be\label{paeq}
C_{1}=3+4\frac{\Lambda}{\gamma^{2}},~C_{2}=\frac{2}{\gamma},
\ee
and the above expressions of the integration constants can be used to relate the observational velocity profiles with the metric parameters.

\subsection{Fitting of the galactic rotation curves}

In this Section we consider the fitting of the  observed rotation curves of a selected sample of  galaxies with the generalized metric of the Weyl geometric gravity, as  given by Eq.~(\ref{meteq}), and which must be considered together with Eq.~(\ref{vtgeq}), giving the tangential velocity of massive test particles.  Since the galactic scale is enormous, and the gravity is essentially weak, except for the region containing supermassive black holes, the non-relativistic approximation of the model, considered in the previous Section, is valid with great accuracy.

There have been a number of studies on the rotation curves of galaxies in the various modified gravity theories,  with or without postulating the existence of dark matter~\cite{Mo1,Panpanich:2018cxo,Mannheim:2010ti,Green:2019cqm,Chardin:2021ahm,Roshan:2021mfc}. Notably, a generalized spherical symmetric metric from dRGT massive gravity, obtained in~\cite{Panpanich:2018cxo}, contained linear and constant terms in the radial coordinate $r$, in addition to the naturally emerging cosmological term proportional to $r^{2}$.  The linear term actually represents the {\it universal acceleration}, originally proposed by Milgrom~\cite{Mo1} in the Modified Newtonian Dynamics~(MOND) model, and it can fit generically, and almost universally, with various kinds of galactic rotation curves, as shown in~\cite{Panpanich:2018cxo}.

The metric function in the Weyl geometric gravity, given by Eq.~(\ref{meteq}), for $\delta=0$ can similarly be cast in the generalized form,
\be\label{meteq1}
e^\nu=1 -\frac{r_g}{r}+\gamma r-\frac{\Lambda r^2}{3}.
\ee

The new parameters $\gamma $ and $\Lambda$, determined by the scalar field profile parameters $C_{1}$ and $C_{2}$, and the constant $C_{3}$, can be used to address the dark matter and the dark energy problems in a single framework.  Interestingly enough, it will be shown later that $C_{1}\sim 10^{3}-10^{4}$~(given by Eq.~(\ref{paeq})) matches the observed cosmological constant value, and the best-fit values of $\gamma$ are relatively constant for all galaxies.

In contrast to the dRGT massive graviton model \cite{Panpanich:2018cxo},  where the graviton mass is related to both $\gamma$ and $\Lambda$, in the Weyl geometric gravity theory $\gamma$ is determined by the scalar field profile, while $\Lambda$ is a free parameter~(since $C_{3}$ is an integration constant). We can thus fix $\Lambda = 1.11247\times 10^{-52}$ m$^{-2}$, as obtained  from the dark energy density of the Universe~\cite{Ade:2015xua}. Then, we leave only one free parameter $\gamma$ of the model to be fit with the observations. However, each galaxy has its own undetermined distribution of baryonic visible masses in the form of bulge~(including the central black hole), disk, and gas components, respectively,  which will be parameterised by $x,y,z$ in the form
\be
\frac{GM(r)}{r} = x v_{\rm b}^{2} + y v_{\rm d}^{2} + z v_{\rm g}^{2},  \label{Newgrav}
\ee
where $v_{\rm b}^{2}, v_{\rm d}^{2}, v_{\rm g}^{2}$ are the velocity-square contributions from the bulge, disk, and gas components in the galaxy, respectively.  Another crucial difference from the dRGT model is that in the Weyl geometric gravity theory there is no additional energy form, such as the massive graviton profile. The mass function $M(r)$ has only contribution from the visible baryonic matter, and there is no negative energy density. We will present the fittings of the Weyl geometric gravity theory for two different kinds of representative galaxies; for spiral galaxies,  including the Milky Way, and for LSB~(Low Surface Brightness) galaxies, respectively. We have also found that most galaxies in the data set~\cite{Lelli:2016zqa,Sofue:2008wt,deBlok:2002vgq,KuziodeNaray:2007qi,vizier} can be fitted in a similar way.

Galaxies normally have visible mass distributions up to large values of $r$. Therefore, the metric at $r$ is not exactly given by the vacuum metric Eq.~(\ref{meteq1}), but by a metric that looks more like the interior of a star.  However, since the pressure of the visible matter in the galaxy is very small, the relativistic effects of the pressure of the baryonic matter are negligible. Hence, the relations (\ref{vtgeq}), (\ref{meteq1})  and (\ref{Newgrav}) can then be used as a very good approximation of the full Tolman-Oppenheimer-Volkoff~(TOV) equation. The velocity formula is then
\be
v^{2}=\displaystyle{\frac{\frac{GM(r)}{r}+\frac{\gamma  r}{2}-\frac{1}{3} \Lambda  r^2}{1-\frac{2GM(r)}{r}+\gamma r -\frac{1}{3} \Lambda  r^2}},  \label{veq}
\ee
where $M(r)/r$ is given by Eq.~(\ref{Newgrav}).

In our rotation-curve fittings, given in Table~\ref{tfit}, we use the data sets from ~\cite{Lelli:2016zqa,Sofue:2008wt,deBlok:2002vgq,KuziodeNaray:2007qi,vizier}. A few examples of the comparisons of the fitting plots and of the observational data are given in the Appendix~\ref{app}.

\begin{table*}[htbp]
 \begin{tabular}{|c||c|c|c|c|}
\hline
& & & & \\[-.4em]
Galaxy &$\gamma =\displaystyle{\frac{2}{C_{2}}}~(10^{-28}~{\rm m}^{-1})$&$(x,y,z)$ &$C_{1}~(10^{4})$& $\left(\chi^{2}_{\rm red}, N-f\right)$
\\[.5em]
\hline
\hline
& & & & \\[-.4em]
Milky Way&$5.2492$&$(1,1,0)$ & $0.1618$&$(5.49, 571)$
\\[1em]
&$5.7490$~(de Vaucouleurs)&$(1,1,0)$ & $0.1349$&$(4.43$~(de Vaucouleurs), 571)
\\[1em]
\hline
& & & & \\[-.4em]
NGC6195~(SAb)&$6.7417$&$(0.7,0.4427,1)$ & $0.0982$&$(1.84, 21)$
\\[1em]
\hline
& & & & \\[-.4em]
NGC6946~(SABcd)&$6.1454$&$(0.4580,0.6127,1)$& $0.1181$&$(2.91, 47)$
\\[1em]
\hline
& & & & \\[-.4em]
NGC2955~(SABb)&$7.2286$&$(0.6992,0.5653,1)$& $0.0855$&$5.36, 21$ \\[1em]
&$5.6224$&$(0.7053,0.5742,2)$ & $0.1411$&$(5.33, 21)$
\\[1em]
\hline
& & & & \\[-.4em]
NGC0024~(SA(s)c)&$5.6301$&$(0,1.8065,1)$& $0.1395$&$(0.33, 27)$ \\[1em]
\hline
& & & & \\[-.4em]
F563-1~(LSB)&$3.6772$&$(0,4.4807,1)$& $0.3294$&$(1.29, 15)$ \\[1em]
\hline
& & & & \\[-.4em]
UGC4173~(LSB)&$1.5114$&$(0,0,1)$& $1.9483$&$(0.14, 12)$
\\[1em]
\hline
& & & & \\[-.4em]
NGC4455~(LSB)&$5.5213$&$(0,0,1)$& $0.1463$&$(0.61, 19)$
\\[1em]
\hline
& & & & \\[-.4em]
UGC3371~(LSB)&$4.9756$&$(0,0,1)$& $0.1800$&$(0.28, 17)$
\\[1em]
\hline
& & & & \\[-.4em]
UGC5005~(LSB)&$2.3908$&$(0,2.9603,1)$& $0.7788$&$(0.20, 9)$
\\[1em]
\hline
& & & & \\[-.4em]
UGC6614~(LSB)&$4.5759$&$(0.5467,0.8821,1)$& $0.2128$&$(0.73, 10)$
\\[1em]
\hline
& & & & \\[-.4em]
DDO189~(LSB)&$2.9057$&$(0,7.0393,1)$& $0.5273$&$(0.15, 9)$
\\[1em]
\hline
& & & & \\[-.4em]
DDO064~(LSB)&$4.2256$&$(0,0.9078,1)$& $0.2495$&$(0.57, 12)$
\\[1em]
\hline
& & & & \\[-.4em]
DDO161~(LSB)&$2.2234$&$(0,0.1207,1)$& $0.9005$&$(2.11, 29)$
\\[1em]
\hline

\end{tabular}
    \caption{The parameter $\gamma$, and the scalar field parameter $C_{1}$~(given by Eq.~(\ref{paeq})) of the Weyl geometric gravity, obtained from fitting the theoretical model with a sample of representative spiral and LSB galaxies. }
\label{tfit}
\end{table*}

\begin{figure*}[htp]
	{{\includegraphics[width=0.9\textwidth]{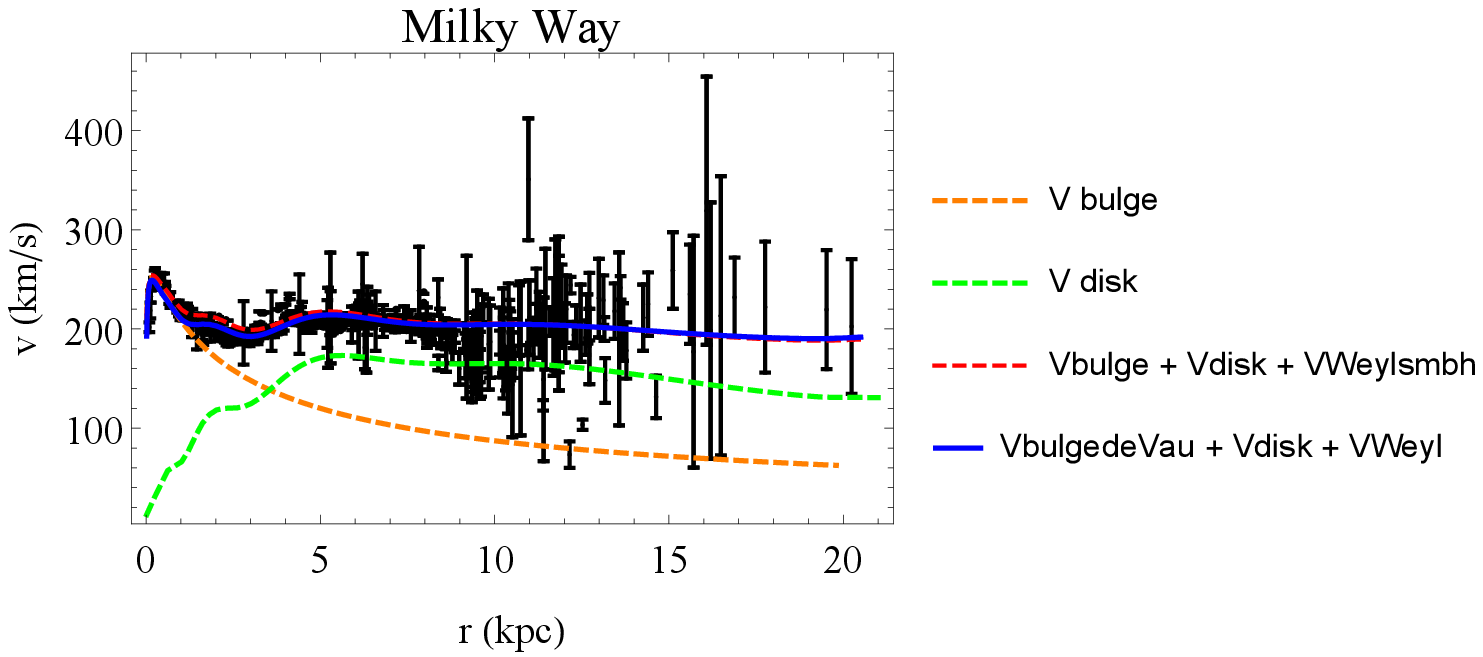}}
	{\includegraphics[width=0.5\textwidth]{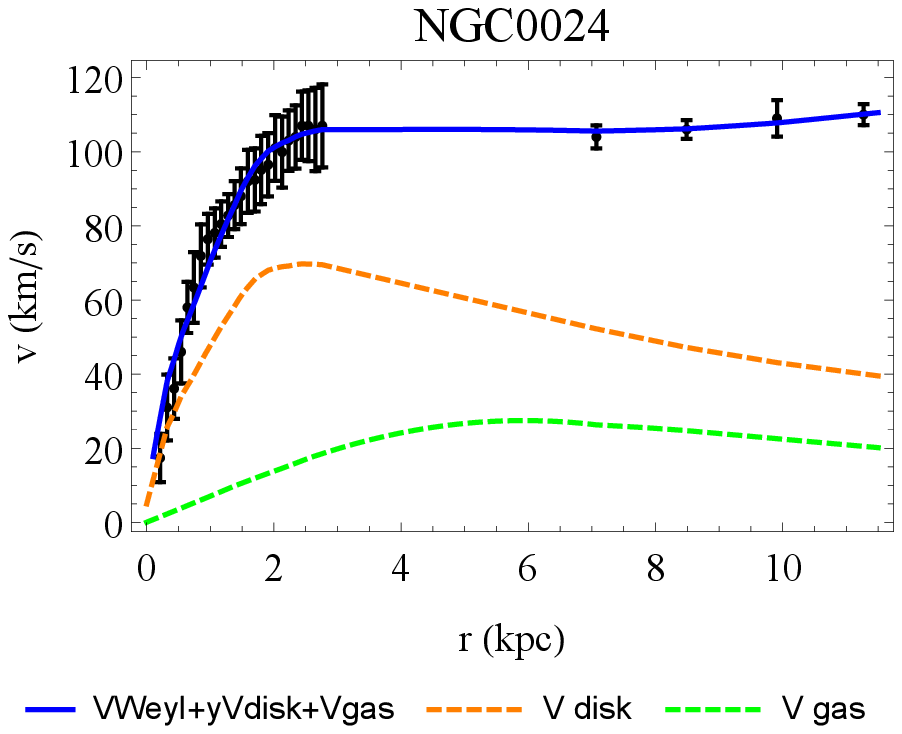}}\hfill
	{\includegraphics[width=0.5\textwidth]{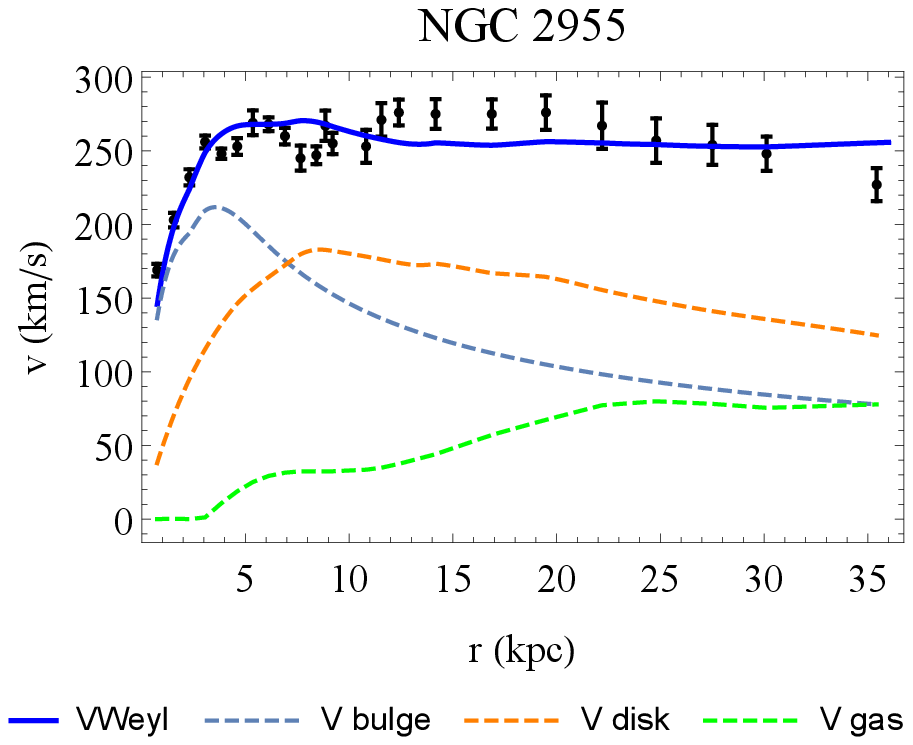}}}\hfill
	\caption{Examples of fittings of the galactic rotation curves of Spiral galaxies. ``VWeyl" represents the contribution from the metric of the Weyl geometric gravity. In the case of the Milky Way the mass contribution from the central supermassive black hole is also contained in the metric. ``VbulgedeVau" represents the refit of the de Vaucouleurs bulge profile found in~\cite{Panpanich:2018cxo}. In the case of the NGC 2955 galaxy,  all the bulge, disk, and gas contributions to the velocity "VWeyl" are included in the metric function of the Weyl geometric gravity. }
	\label{fitspifig}
\end{figure*}

The best-fit values of $\gamma$ in Table~\ref{tfit} are of the order of $10^{-28}$ m$^{-1}$. The goodness-of-fit is measured by the reduced chi square values
\be
\chi^{2}_{\rm red}=\frac{1}{N-f}\sum_{i=1}^{N}\frac{(v_{i}^{\rm th}-v_{i}^{\rm ob})^{2}}{(\sigma_{i}^{\rm ob})^{2}},
\ee
where $N$ is the number of the fitting data points, $f$ is the number of fitting parameters, $\sigma_{i}^{\rm ob}$ is the velocity uncertainty from the data set, and $v_{i}^{\rm th~(ob)}$ are the theoretical~(observed) velocities, respectively. Note that for some data points of the Milky Way data set, the velocity uncertainty $\sigma_{i}^{\rm ob}$ is listed as 0, which we replace by a moderate estimate of $5$ km s$^{-1}$. In the fitting of the Milky Way, we perform two kinds of fittings, one with the supermassive black hole separated from the bulge profile, as given by Ref.~\cite{Sofue:2008wt}, and another with the best-fit of the de Vaucouleurs bulge parameters, given in ~\cite{Panpanich:2018cxo}. It can be seen from the values of $\chi^{2}_{\rm red}$ in Table~\ref{tfit} and from Fig.~\ref{fitspifig} that the Weyl geometric gravity model fits the de Vaucouleurs bulge profile slightly better than the standard dark matter models.

By multiplying the best-fit value of $\gamma$ by $c^{2}$, we obtain a constant acceleration of the order of $10^{-11}$ ms$^{-2}$). The values of $\gamma$ are identical to the similar values of the coefficient of the linear term of the  dRGT model, previously explored in ~\cite{Panpanich:2018cxo}. This result is due to the use of the same fitting metric form. Note that in the fittings with the simplest Weyl geometric gravity model of LSB galaxies, the mass-to-light ratio~($M/L_{*}$) of the disk component could vary by an order of magnitude from the observed values. For example in F563-1, UGC5005, DDO189, the best-fit values of $y$, the disk contribution, are roughly $4.5, 3.0, 7.0$ respectively, while for DDO161, the best-fit of $y$ is only $0.12$. Notably for DDO161, $\chi^{2}_{\rm red}$ of the fitting is relatively large, due to exceptionally small error bars. The $\chi ^2$ for the Milky Way galaxy fit is enlarged by a number of points separated from the theoretical curve, which could be the result of other astrophysical or physical effects.

Also, if we take into account the uncertainty in the radial distance, the values of $\chi^{2}_{\rm red}$ would reduce even further. The same is also true for other spirals, except NGC0024, where the fit is already excellent. However, the core-cusp problem~\cite{deBlok:2009sp} can still be seen in the fitting of the F563-1 galaxy, where the Weyl geometric gravity shows apparent cusp profile, deviating from the observed values in the core region, leading to a relatively large value of $\chi^{2}_{\rm red}$.

Despite the similarity to the dRGT model, the physical interpretation in the case of the Weyl geometric gravity is in terms of the scalar and Weyl vector fields profiles,  parameterised by $C_{1}$ and $C_{2}$.  Our fitting values of $\gamma \sim 10^{-28}$ m$^{-1}$ implies that $C_{2}=2/\gamma \sim 10^{28}$ m is extremely large, even {\it larger than} cosmic horizon scale, which is $3/\sqrt{\Lambda}\sim 10^{26}$ m.  For small $r \ll C_{2}$, the general relativistic approximation becomes very accurate, and the effects of the linear $\gamma r$ term are negligible.  For sufficiently large values of $r$ in the galactic scale, the effect of the linear term becomes significant, as  compared to the Schwarzschild term, leading to the  observed non-decreasing rotation curves of galaxies.

In the Weyl geometric gravity model, each galaxy is the source of this linear term, which extends up to the cluster and supercluster scales, thus governing the distribution and motion of galaxies within.  It would be interesting to explore such effects on the motion of galaxies in the clusters and superclusters in a future work. Our Weyl geometric gravity model also suggests that the cosmological constant $\Lambda = -3C_{3}$ might vary from one galaxy to another, since it is an integration constant in the metric, determined by the field equations, the Weyl scalar and vector fields, and the mass profile of each galaxy. Hence, $C_3$ is not a fundamental constant of the theory, originally inserted in the action (\ref{a3}).

However, the universal smallness of $\Lambda$ on the cosmological scale is not explained in this simplest model of conformally invariant and explicit conformal symmetry breaking Weyl geometric gravity.

\section{Discussions and final remarks}\label{sect5}

After many decades of intensive investigations, a convincing physical explanation of the dynamics, and properties, of the galactic rotation curves is still missing. The very existence, and the nature of dark matter, usually proposed to explain the astrophysical observations at the galactic scale, still represents a major challenge for contemporary physics. There are a number of intriguing questions related to the rotation curves,  like, for example, their universal behavior, or the very close correlation between dark matter, and the baryonic component in the galaxy. To explain the observational facts, the most commonly used models are either based on particle physics, and Newtonian gravity-thus assuming the existence of a mysterious dark matter particle,  or on modified gravity theory, representing extensions of general relativity.

In the present paper we have followed the second line of research, as mentioned above, by considering, and developing, an alternative geometric point of view to the dark matter problem, by considering  the possibility that the behavior of the galactic rotation curves can naturally be explained by conformally invariant gravitational theories constructed in a Weyl geometry, in which two additional degrees of freedom, a scalar and a vector one, do naturally appear. Beginning with the simplest possible action in Weyl geometry, constructed from the square of the Weyl scalar and the strength of the Weyl vector, by introducing an auxiliary scalar field, an equivalent gravitational action theory is obtained in the Riemann geometry, which is linear in the Ricci scalar, and with the scalar field nonminimally coupled to $R$ and $\omega _{\mu}$. We have considered the static spherically symmetric solutions of the theory for a specific choice of the Weyl vector, having only a nonzero radial component. This choice has the advantage of allowing to obtain an exact analytic solution of the field equations, which can be effectively used to test the model.  The gravitational field equations also contain some extra, scalar field dependent terms, which can be interpreted as corresponding to an effective geometric mass (and density), which is a direct consequence of the Weyl geometry existing in the vacuum outside the galactic baryonic matter distribution, and which can explain the observed dynamical motion of the galactic rotation curves. We have investigated in detail the behavior of the tangential velocity of massive test particles in Weyl geometric gravity, and, after obtaining the general formalism, some simple cases were also investigated.

One of the fundamental aspects of any physical theory is its comparison with observations. In the present work we have also compared the theoretical predictions of Weyl geometric gravity in spherical symmetry with a selected sample of galactic rotation curves data. In order to perform a proper fit the effects of the luminous matter in the galaxy must also be considered. However, in the presence of baryonic matter, the conformal symmetry in Weyl geometric gravity theory will be broken, either explicitly, or spontaneously.  With the ansatz (\ref{conbrL}), the general metric function (\ref{meteq}) of the conformally broken Weyl geometric gravity naturally contains in the general relativistic limit a linear term in the radial coordinate, as well as a quadratic term in $r$, corresponding to the presence of the cosmological constant.  This generalized spherically symmetric metric can be fitted to the rotation curves of spiral and LSB galaxies, with fitting parameters given, for a selected set of representative galaxies, in Table~\ref{tfit}.

The linear term in the metric plays the role of a universal acceleration, which is of the same order of magnitude for all galaxies, regardless of their various shapes and types.  Since the linear term is defined with respect to the galactic centers, with each galaxy having its own Weyl scalar field profile, it can potentially explain the gravitational lensing results of colliding galaxies, like, for example, in the Bullet Cluster. The center of the dark matter halo for each colliding galaxy can be interpreted as the center of the conformally broken Weyl spacetime, coupled to the Weyl scalar of each galaxy.  It would be interesting to perform a detailed quantitative analysis of the weak lensing in colliding galaxies in Weyl geometric gravity in a  future work.

Weyl geometric gravity naturally leads to an effective geometric ``dark matter" mass $M_{DM}$, and to an associated effective density $\rho_{DM}$. It is also possible, at least theoretically, to observationally constrain $M_{DM}$ and $\rho _{DM}$ by using a method based on Jeans equation \cite{BT08}. In order to do so we first assume that each galaxy contains a single stellar population, assumed to be pressure supported, and which is in a dynamical equilibrium. Moreover, the galactic system also contains the gravitational
contribution resulting from the presence of the Weyl geometric gravity, and of the modification of the space-time structure due to the presence of Weyl geometry. By assuming static spherical symmetry, the effective geometric mass profile induced  by the Weyl geometric gravity effects, which, from an astrophysical point of view, represent the mass profile of the ``dark matter" halo, are related to the moments of the stellar distribution function by the Jeans equation \cite{BT08}
\begin{equation}\label{eqf}
\frac{d}{dr}\left[ \rho _{s}\left\langle v_{r}^{2}\right\rangle \right] +%
\frac{2\beta \rho _{s}\left( r\right) \left\langle v_{r}^{2}\right\rangle }{r}=-\frac{G\rho _{s}M_{DM}(r)}{r^{2}},
\end{equation}%
where $\rho _{s}(r)$, $\left\langle v_{r}^{2}\right\rangle $, and $\beta
(r)=1-\left\langle v_{\theta }^{2}\right\rangle /\left\langle
v_{r}^{2}\right\rangle $ denote the three-dimensional stellar density, the radial
velocity dispersion of the stars, and the orbital anisotropy of the stellar component, respectively. In Eq.~(\ref{eqf})  $\left\langle v_{\theta }^{2}\right\rangle$ denotes the dispersion of the tangential velocity.

Under the assumption  that the anisotropy $\beta $ is constant, the Jeans equation (\ref{eqf}) has the general
solution \cite{MaLo05}
\begin{equation}\label{108}
\rho _{s}\left\langle v_{r}^{2}\right\rangle  =Gr^{-2\beta
}\int_{r}^{\infty }s^{2(1-\beta )}\rho _{s}\left( s\right) M_{DM}\left(
s\right) ds.
\end{equation}

One can now use the various mass profiles obtained in the different models of Weyl geometric gravity to relate the model parameters to the stellar densities, and velocity dispersions. This can be done by projecting Eq.~(\ref{108}) along the line of sight, and relating the ``dark matter" mass profile to two observable astrophysical profiles, the projected stellar density $I(R)$, and to $\sigma _p(R)$, the stellar velocity dispersion, respectively, according to the relation \cite{BT08}
\begin{equation}
\sigma _{P}^{2}(R)=\frac{2}{I(R)}\int_{R}^{\infty }\left( 1-\beta \frac{R^{2}%
}{r^{2}}\right) \frac{\rho _{s}\left\langle v_{r}^{2}\right\rangle r}{\sqrt{%
r^{2}-R^{2}}}dr.
\end{equation}

Once a projected stellar density model $I(R)$ is given, one can obtain the
three-dimensional stellar density by using the equation \cite{BT08}
\be
\rho _{s}(r)=-(1/\pi
)\int_{r}^{\infty }\left(\frac{ dI}{dR}\right) \left( R^{2}-r^{2}\right) ^{-1/2}dR.
\ee

Hence, if the galactic metric, the stellar velocity dispersion $\left\langle v_{r}^{2}\right\rangle$, and  the stellar density profile $I(R)$, respectively,  are known, from the integral equation Eq.~(\ref{108}) one can obtain the explicit form of the Weyl geometric dark matter mass profile $M_{DM}$.

One of the simplest analytic projected density profiles is represented by the Plummer profile \cite{BT08},
\be
I(R)={\cal L}_0\left(\pi r_{half}^2\right)^{-1}\left(1+\frac{R^2}{r_{h}^2}\right)^{-2},
\ee
where ${\cal L}_0$ denotes the total galactic luminosity, while $r_{h}$ represents the projected half-light radius, giving the radius of the cylinder containing half of the total galactic luminosity. Hence, with the use of the Plummer profile some constraints on the equivalent geometric dark mass distribution in Weyl geometric gravity can be obtained as follows. By using the Plummer profile we obtain for the stellar density the expression
\be
\rho _s(r)=\frac{3 {\cal L}_0}{4 \pi r_{half}^3
   \left(1+r^2/r_{h}^2\right)^{5/2}}.
\ee

By adopting for the dark matter mass profile Eq.~(\ref{91}), corresponding to the choice of the Weyl vector as $\omega _1=1/\alpha r$, in the first order of approximation for the stellar velocity dispersion we obtain the expression
\begin{widetext}
\bea
\left\langle v_{r}^{2}\right\rangle&\approx& \frac{G(rr_{h})^{-2\beta }}{48\sqrt{\pi }}\Bigg\{ \frac{3c_{1}\Gamma \left(
-\beta -\frac{1}{4}\right) \Gamma \left( \beta +\frac{3}{4}\right) \left[
(24\beta -26)r^{4}+5r_{h}^{2}(8(1-2\beta )\beta +3)r^{2}+2(8(1-2\beta )\beta
+3)r_{h}^{4}\right] }{\sqrt{r_{h}}}\nonumber\\
&&+\frac{4r_{h}\Gamma \left( \frac{1}{2}%
-\beta \right) \Gamma (\beta )\left[ (11-6\beta )r^{4}+5(4(\beta -2)\beta
+3)r^{2}r_{h}^{2}+2(4(\beta -2)\beta +3)r_{h}^{4}\right] }{\alpha }\nonumber\\
&&+72\Gamma
(-\beta )\Gamma \left( \beta +\frac{1}{2}\right) \left[ (3\beta
-4)r^{4}-10(\beta -1)\beta r^{2}r_{h}^{2}-4(\beta -1)\beta r_{h}^{4}\right]
\Bigg\},
\eea
\end{widetext}
where $\Gamma (z)$ denotes the Euler gamma function, $\Gamma (z)=\int_0^{\infty}{t^{z-1}e^{-t}dt}$. Hence, the coupling parameter $\alpha$ in the expression of the Weyl vector, can be determined, at least in principle, from the study of the stellar velocity dispersion in the galaxy.

From both observational and theoretical point of view an important problem is to obtain an estimate of the upper bound of the cutoff radius of the constancy of the tangential velocities. Even that in principle in the Weyl geometric gravity theory in the considered models the rotation curves do extend to infinity, this cannot be the case in a realistic astrophysical environment. Hence, in the more general models  one should find expressions of the tangential velocity that in the large $r$ limit decay to zero. For example, if the exact functional form of $\Phi$ is known, the numerical value of $r$ at which the rotational velocity decreases to zero can be accurately obtained.

On the other hand, we need to mention that from a physical point of view the increase to $c$ of the galactic rotation curves in the exact solution of the Weyl geometric gravity
happens only when we reach far beyond the cosmic horizon scale, which is not a physical situation.

Moreover, even though in the present theoretical approach  we consider
only one galactic center to generate the metric, and the rotational velocity profile, in realistic astrophysical situations we need to take into account the gravitational effects
of other galaxies, whose scalar or vector mass profiles also influence the speed of motion of the particle in the galactic halo. When we consider the limits of large distances,
the contributions of the other galaxies become important, and the resulting total speed will be normalized in the cluster and supercluster scale.  In order to estimate the total velocity of a particle in the galactic halo, we need to also consider the effects of the neighboring galaxies clustering together. Only such a full investigation of the galactic dynamics would allow to obtain an exact quantitative answer on the asymptotic behavior of the rotational velocities.

We would like also to point out ar this moment that in order to explain the phenomenology related to the dynamic of galactic rotation curves one needs to obtain, from a theoretical model, a force that scales as $1/r$ instead of the usual Newtonian force, scaling as $1/r^2$. A similar approach can also be applied to the case of the galaxy clusters. However, it is more difficult to explain simultaneously the properties of individual galaxies, and of the galaxy clusters. For example, if one uses the MOND  fits \cite{Mo1} that reproduce well the properties of the galaxies, in order to explain the galactic cluster properties some dark matter is needed, usually assumed to be in the form of massive neutrinos. Dark matter has also important implications for cosmological perturbations, and structure formation, and these aspects are more difficult to explain by using modified gravity type approaches, including the MOND type ones. Hence,  in order to estimate the possible implications of a modified gravity model for dark matter one must investigate not  only galactic phenomenology, but also galactic clusters, and, more importantly, its cosmological implications.

It is also important to mention that our investigations have been restricted to a very limited class of Weyl geometric gravity models, for which exact solutions can be obtained. In the general case the expressions of the tangential velocities, as well as the fitting with the observational data must be performed by numerically solving the gravitational field equations, without specifying a priori the form of the scalar field, or of the Weyl vector. Such an analysis would provide deeper insights into the problem of dark matter in Weyl geometric gravity, and in its role in galactic dynamics.

However, despite their simplicity, the  models considered in the present paper allow to express al the relevant geometrical and physical quantities, including the scalar field, the Weyl vector, the effective geometric mass, playing the role of the dark matter, and its corresponding effective geometric density profile, in terms of observable
parameters, like the tangential velocity of test particles, and the parameters of the baryonic matter. Therefore, these results open the possibility of directly constraining the Weyl geometric gravity theory  by using direct astrophysical and astronomical observations at the galactic and extra-galactic scale. In this work,  we have presented some fundamental theoretical tools necessary for a comprehensive comparison of the predictions of the Weyl geometric gravity theory with the observational results.

\section*{Acknowledgments}

The work of TH is supported by a grant of the Romanian Ministry of Education and Research, CNCS-UEFISCDI, project number PN-III-P4-ID-PCE-2020-2255 (PNCDI III).

\appendix

\section{Sample Plots of Rotation Curves Fitting in Spiral and LSB Galaxies} \label{app}

In the present Appendix we present the results of the fitting of a small selected sample of galactic rotation curves with the theoretical expression of the rotational velocity in Weyl geometric gravity. We have selected a total of seven galaxies, including the Milky Way, two spiral galaxies,  NGC0024 and NGC2955, respectively,
and four LSB galaxies, UGC6614, F563-1, DDO064, and DDO161, respectively. The values of the parameters of the Weyl geometric gravity model are presented in Table~\ref{tfit}.

\begin{figure*}[htp]
      {{\includegraphics[width=0.45\textwidth]{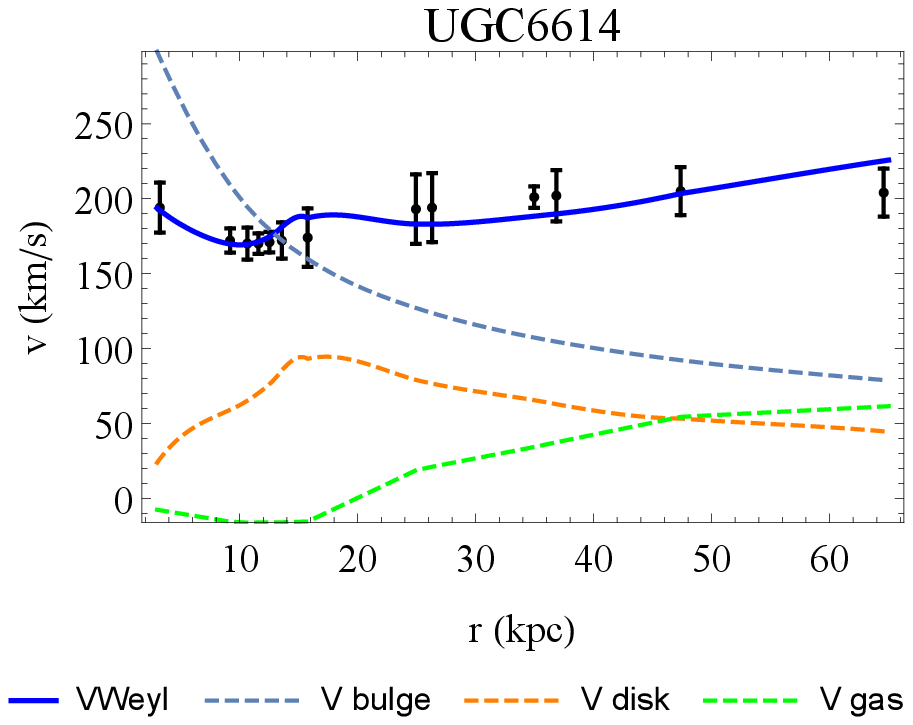}}\hfill
	{\includegraphics[width=0.45\textwidth]{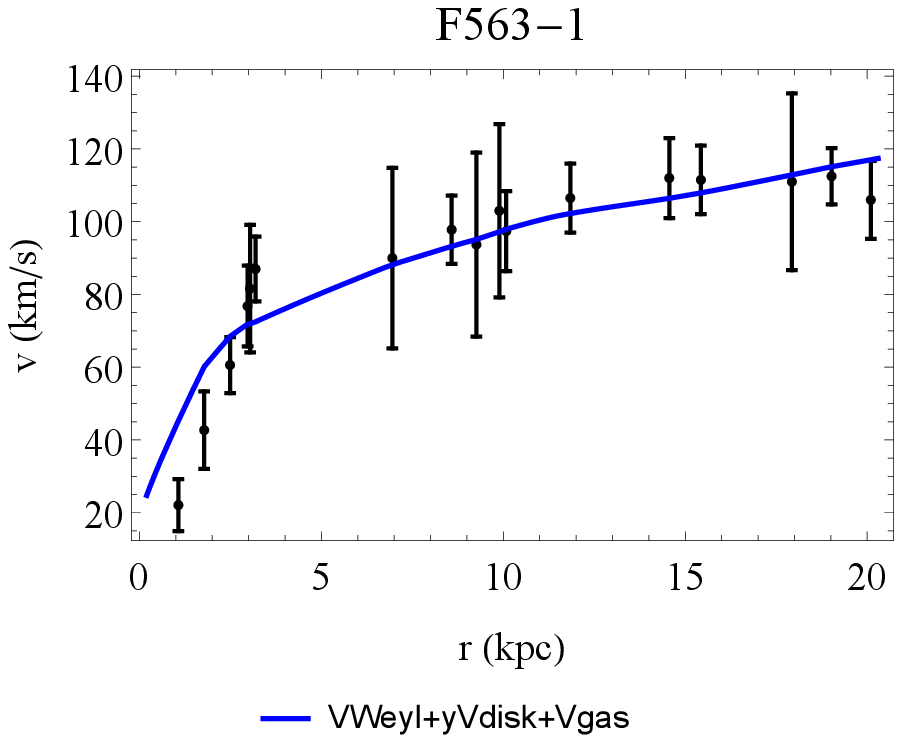}}}\hfill
      {{\includegraphics[width=0.45\textwidth]{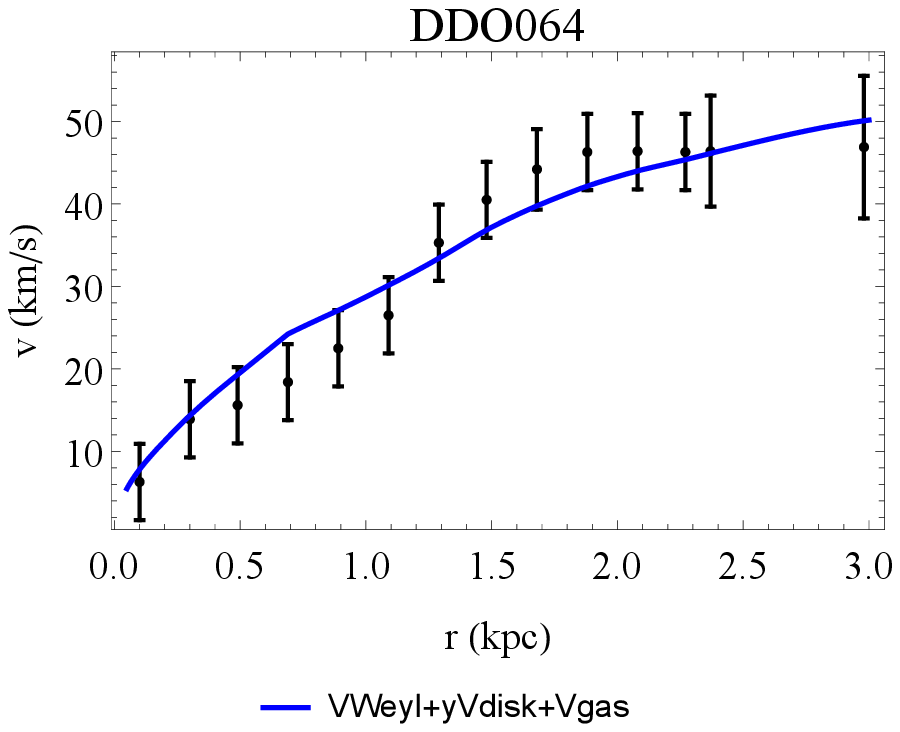}}\hfill
	{\includegraphics[width=0.45\textwidth]{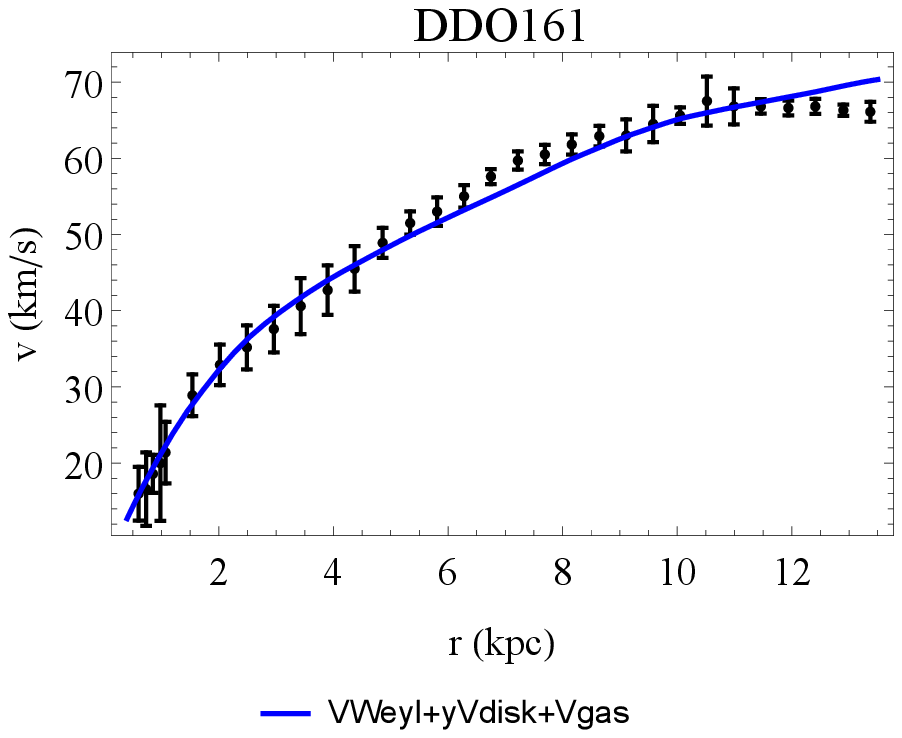}}}\hfill
	\caption{Examples of fittings of the rotation curves of LSB galaxies. "yVdisk" denotes the best fit of the disk contribution parameter $y$, which is generally not 0.  The corresponding  values of $y$ are listed in Table~\ref{tfit}.}
	\label{fitLSBfig}
\end{figure*}


\clearpage

\begin{thebibliography}{99}

\bibitem{Ein} A. Einstein, Sitzungsberichte der Preussischen
Akademie der Wissenschaften zu Berlin, {\bf 844}, (1915).

\bibitem{Hilb} D. Hilbert, Nachrichten von der Gesellschaft der Wissenschaften zu G\"{o}ttingen - Mathematisch - Physikalische Klasse {\bf 3}, 395 (1915).

\bibitem{Riemm} B, Riemann, Ueber die Hypothesen, welche der Geometrie zu Grunde liegen, Abhandlungen der K\"{o}niglichen Gesellschaft der Wissenschaften zu G\"{o}ttingen, {\bf 13}, 1 (1867).

\bibitem{Ric} G. Ricci and T. Levi-Civita,  Mathematische Annalen  {\bf 54},  125 (1900).

\bibitem{Will} C. Will, Living Rev. Relativity {\bf 17},  4 (2014).

\bibitem{GW1}  B. P. Abbott, et al. (LIGO Scientific Collaboration and Virgo Collaboration), Phys. Rev. Lett. {\bf 116}, 061102 (2016).

\bibitem{GW2} B. P. Abbott, et al. (LIGO Scientific Collaboration and Virgo Collaboration), Phys. Rev. Lett. {\bf 116},  241102 (2016).


\bibitem{AU1}  A. G. Riess et al., Astron. J. {\bf 116}, 1009 (1998).
\bibitem{AU2} S. Perlmutter et al., Astrophys. J. {\bf 517}, 565 (1999).
\bibitem{AU3} P. de Bernardis et al., Nature {\bf 404}, 955 (2000).
\bibitem{AU4} S. Hanany et al., Astrophys. J. {\bf 545}, L5 (2000).
\bibitem{AU5} R. A. Knop et al., Astrophys. J. {\bf 598}, 102 (2003).
\bibitem{AU6} M. Hicken, W. M. Wood-Vasey, S. Blondin, P. Challis, S. Jha, P. L. Kelly, A. Rest, and R. P. Kirshner,
Astrophys. J. {\bf 700}, 1097 (2009).
\bibitem{AU7}  R. Amanullah et al., Astrophys. J. {\bf 716}, 712 (2010).

\bibitem{Pl} N. Aghanim et al. Planck Collaboration, Astronomy \& Astrophysics {\bf 641}, A6 (2020)
\bibitem{BAO1} K. S. Dawson et al., 2013, Astron. J. {\bf 145}, 10 (2013).
\bibitem{BAO2}K. S. Dawson et al., Astron. J. {\bf 151}, 44 (2016).
\bibitem{BAO3} M. Gatti et al., Monthly Notices of the Royal Astronomical Society {\bf 510}, 1223 (2022).
\bibitem{Wein} D. H. Weinberg, M. J. Mortonson, D. J. Eisenstein, C. Hirata, A. G. Riess, and E. Rozo, Physics Reports {\bf 530}, 87 (2013).
\bibitem{Einl} A. Einstein, Sitzungsberichte der K\"{o}niglich Preussischen Akademie der Wissenschaften, Berlin, part 1: 142 (1917).

\bibitem{DE1} A. Joyce, B. Jain, J. Khoury, and M. Trodden, Phys. Rept. {\bf 568}, 1 (2015).

\bibitem{DE2}  A. Joyce, L. Lombriser, and F. Schmidt, Annu. Rev.
Nucl. Part. Sci. 66, 95 (2016).
\bibitem{DE3} A. N. Tawfik and E. A. El Dahab, Gravitation and Cosmology {\bf 25}, 103 (2019).
\bibitem{DE4} N. Frusciante and L. Perenon, Phys. Rept. {\bf 857}, 1 (2020).

\bibitem{DM1} A. Arbey and F. Mahmoudi, Progress in Particle and Nuclear Physics {\bf 119}, 103865 (2021).
\bibitem{DM2} E. Oks, New Astronomy Reviews {\bf 93}, 101632 (2021).
\bibitem{DM3} J. de Dios Zornoza, Universe {\bf 7}, 415 (2021).
\bibitem{RC1} P. Salucci, C. Frigerio Martins, and A. Lapi, arXiv:1102.1184 (2011).
\bibitem{RC2} J. Binney and S. Tremaine, Galactic dynamics, Princeton University Press, Princeton (1987).
\bibitem{RC3} M. Persic, P. Salucci, and F. Stel, Mon. Not. R. Astron. Soc. {\bf 281}, 27 (1996).
\bibitem{RC4} A. Boriello and P. Salucci, Mon. Not. R. Astron. Soc. {\bf 323}, 285 (2001).
\bibitem{RDM1} J. M. Overduin and P. S. Wesson, Phys. Repts. {\bf 402}, 267 (2004).
\bibitem{RDM2} V. Beylin, M. Khlopov, V. Kuksa, and N. Volchanskiy, Universe {\bf 6}, 196 (2020).
\bibitem{RDM3} O. Lebedev, Progress in Particle and Nuclear Physics {\bf 120}, 103881 (2021).
\bibitem{RDM4} L. Bian, X. Liu, and K.-P. Xie, Journal of High Energy Physics {\bf 2021}, 175 (2021).

\bibitem{B1} C. G. Boehmer  and T. Harko, 	JCAP {\bf 0706}, 025 (2007).

\bibitem{B2} T. Harko, JCAP {\bf 1105}, 022 (2011).

\bibitem{B3} M. Craciun and T. Harko, Eur. Phys. J. {\bf C 80}, 735 (2021).

\bibitem{Mo1} M. Milgrom, Astrophys. J. {\bf 270}, 365 (1983).

\bibitem{Boh} C. G. Boehmer, T. Harko, and F. S. N. Lobo, Astropart. Phys. {\bf 29}, 386 (2008).

\bibitem{HMP}  T. Harko, T. S. Koivisto, F. S. N. Lobo and G. J. Olmo, Phys. Rev. {\bf D 85}, 084016 (2012).

\bibitem{HMP1} S. Capozziello, T. Harko, T. S. Koivisto, F. S. N. Lobo, and G. J. Olmo, JCAP {\bf 07}, 024 (2013).

\bibitem{HMP2} S. Capozziello, T. Harko, T. S. Koivisto, F. S. N. Lobo, and G. J. Olmo, Astroparticle Physics {\bf 50}, 65 (2013).

\bibitem{MG1}  O. Bertolami, C. G. Boehmer, T. Harko, and F. S. N. Lobo, Phys. Rev. {\bf D 75}, 104016 (2007).
\bibitem{MG2} T. Harko, F. S. N. Lobo, S. Nojiri, and S. D. Odintsov, Phys. Rev. {\bf D 84}, 024020 (2011).
\bibitem{MG3} T. Harko and F. S. N Lobo, Eur. Phys. J. {\bf C 70}, 373 (2010).
\bibitem{MG4} T. Harko and F. S. N. Lobo, Galaxies 2014, 410 (2014).
\bibitem{MG5} T. Harko and F. S. N. Lobo, Int. J. Mod. Phys. {\bf D 21}, 1242019 (2012).
 \bibitem{MG6} T. Harko and F. S. N. Lobo, Int. J. Mod. Phys. {\bf D 29}, 2030008 (2020).
\bibitem{MG7} Z. Haghani and T. Harko, Eur. Phys. J. {\bf C 81}, 615 (2021).
\bibitem{MG8} T. Harko, N. Myrzakulov, R. Myrzakulov, and S. Shahidi, Physics of the Dark Universe {\bf 34}, 100886 (2021).

\bibitem{MDM1} T. Katsuragawa and S. Matsuzaki, Physical Review {\bf D 95}, 044040 (2017).

\bibitem{MDM2} B. L'Huillier, H. A. Winther, D. F. Mota, C. Park, and J. Kim, Mon. Not. Roy. Astron. Soc. {\bf 468}, 3174 (2017).

\bibitem{MDM3} J. W. Moffat and V. T. Toth, Mon. Not. Roy. Astron. Soc. Letters {\bf 482},  L1 (2019).

\bibitem{MDM4} M. Lisanti, M. Moschella, N. J. Outmezguine, and O. Slone, Phys. Rev. {\bf D 100}, 083009 (2019).

\bibitem{MDM5} P. S. Corasaniti, C. Giocoli, and M. Baldi, Phys. Rev. {\bf D 102}, 043501 (2020).

\bibitem{MDM6} A. R. Khalifeh and R. Jimenez, Mon. Not. Roy. Astron. Soc. {\bf 501}, 254 (2021).

\bibitem{MDM7} A. De Felice and S. Mukohyama, JCAP {\bf 2021}, 018 (2021).

\bibitem{MDM8} M. M. Brouwer et al., Astron.\& Astrophys. {\bf  650}, A113 (2021).

\bibitem{book}  T. Harko and F. S. N. Lobo, Extensions of $f(R)$ Gravity: Curvature-Matter Couplings and Hybrid Metric Palatini Theory, Cambridge University Press, Cambridge, UK, 2018

\bibitem{Reuter} M. Reuter and H. Weyer, Phys. Rev. {\bf D 70}, 124028 (2004).

\bibitem{Weyl1} H. Weyl, Sitzungsberichte der K\"{o}niglich Preussischen Akademie der Wissenschaften zu Berlin 1918, 465 (1918).

\bibitem{Weyl2} H. Weyl, Space, Time, Matter, Dover, New York, 1952

\bibitem{Weyl3}  E. Scholz, arXiv:1703.03187 (2017).

\bibitem{Di1} P. A. M. Dirac, Proceedings Royal Society London {\bf A 333}, 403 (1973).

\bibitem{Di2} P. A. M. Dirac, Proceedings Royal Society London {\bf A 338}, 439 (1974).

\bibitem{Di3} N. Rosen, Foundations of Physics {\bf 12}, 213 (1982).

\bibitem{Di4} M. Israelit, Gen. Relativ. Gravit. {\bf 43}, 751 (2011).

\bibitem{Ut1}  R. Utiyama, Progress of Theoretical Physics {\bf 50}, 2080 (1973).
\bibitem{Ut2} R. Utiyama, Progress of Theoretical Physics {\bf 53}, 565 (1975).
\bibitem{Ni} M. Nishioka, Fortsch. Phys. {\bf 33}, 241 (1985).


\bibitem{M1} P. D. Mannheim and D. Kazanas, Astrophys. J. {\bf 342},
635 (1989).

\bibitem{M2} P. D. Mannheim, Found. Phys. {\bf 24}, 487 (1994).

\bibitem{M3} P. D. Mannheim, Found. Phys. {\bf 26}, 1683 (1996).

\bibitem{M4} P. D. Mannheim, Found. Phys. {\bf 30}, 709 (2000).

\bibitem{M5} P. D. Mannheim, Found. Phys. {\bf 37}, 532 (2007).

\bibitem{M6}  P. D. Mannheim, Found. Phys. {\bf 42}, 388 (2012).

\bibitem{P1} R. Penrose, Cycles of Time: An Extraordinary New View of the Universe, Bodley Head, London, UK, 2010

\bibitem{P1a} V. G. Gurzadyan and R. Penrose, Eur.Phys.J. Plus {\bf 128}, 22 (2013).
\bibitem{P2} I. Bars, P. J. Steinhardt, and N. Turok, Phys. Lett. {\bf B 726}, 50 (2013).
\bibitem{P3} R. Penrose, Foundations of Physics {\bf 44}, 873 (2014).
\bibitem{P4} P. Tod, General Relativity and Gravitation {\bf 47}, 17 (2015).
\bibitem{P5} A. Araujo, H. Jennen, J. G. Pereira, A. C. Sampson, and L. L. Savi, Gen. Relativ. Grav. {\bf 47}, 151 (2015).
\bibitem{P6} U. Camara da Silva, A. L. Alves Lima, and G. M. Sotkov, Journal of High Energy Physics {\bf 2016}, 90 (2016).
\bibitem{P7} P. Nurowski, Classical and Quantum Gravity {\bf 38}, 145004 (2021).
\bibitem{H1} G. ’t Hooft, International Journal of Modern Physics {\bf D 24}, 1543001 (2015).

\bibitem{H2} G. ’t Hooft, arXiv:1511.04427 (2015).

\bibitem{Q1} J. M. Nester and H.-J. Yo, Chinese Journal of Physics {\bf 37}, 113 (1999).

\bibitem{Q2} J. Beltran Jimenez, L. Heisenberg, and T. Koivisto, Phys. Rev. {\bf D 98}, 044048 (2018).

\bibitem{Q3} J. Beltran Jimenez and T. S. Koivisto, Phys. Lett. {\bf B 756}, 400 (2016).

\bibitem{Q4} A. Golovnev, T. Koivisto and M. Sandstad, Class. Quant. Grav. {\bf 34}, 145013 (2017).

\bibitem{Q5} Y. Xu, T. Harko, S. Shahidi, and S.-D. Liang, Eur. Phys. J. {\bf C 80}, 449 (2020).

\bibitem{Q6} J.-Z. Yang, S. Shahidi, T. Harko, and S.-D. Liang, Eur. Phys. J. {\bf C 81}, 111 (2021).


\bibitem{Gh1} D. M. Ghilencea, JHEP {\bf 03},  049 (2019).

\bibitem{Gh2} D. M. Ghilencea and H. M. Lee, Phys. Rev. {\bf D 99}, 115007 (2019).

\bibitem{Gh3} D. M. Ghilencea, JHEP {\bf 10}, 209 (2019).

\bibitem{Gh4} D. M. Ghilencea, Phys. Rev. {\bf D 101}, 045010 (2020).

\bibitem{Gh5} D. M. Ghilencea, Eur. Phys. J. {\bf C 80}, 1147 (2020).

\bibitem{Gh6} D. M. Ghilencea, Eur. Phys. J. {\bf C 81}, 510 (2021).

\bibitem{Gh7} D. M. Ghilencea and T. Harko, arXiv:2110.07056 [gr-qc] (2021).

\bibitem{Gh8} D. M. Ghilencea, Eur. Phys. J. {\bf C 82}, 23 (2022).

\bibitem{Gh9} D. M. Ghilencea, arXiv:2203.05381 (2022).

\bibitem{Gh10} Matthias Weisswange, D. M. Ghilencea, and D. St\"{o}ckinger, arXiv:2208.01293 [hep-ph] (2022).

\bibitem{Gh11} T. Harko and S. Shahidi,  Eur. Phys. J. {\bf C 82}, 219 (2022).

\bibitem{Gh12} T. Harko and S. Shahidi, Eur. Phys. J. {\bf C 82}, 1003 (2022).

\bibitem{Gh13} J.-Z. Yang, S. Shahidi, and T. Harko,  Eur. Phys. J. {\bf C 82}, 1171 (2022).

\bibitem{LaLi} L. D. Landau and E. M. Lifshitz, The Classical Field Theory,
Pergamon Press, New York, 1975

\bibitem{Bera} V. A. Berezin, V. I. Dokuchaev, Y. N. Eroshenko, and A. L. Smirnov, JCAP {\bf 11}, 053 (2021).

\bibitem{Berb} V. A. Berezin and V. I. Dokuchaev, Physics {\bf 3}, 814 (2021).

\bibitem{Ma03a}  T. Matos and F. S. Guzman, Class. Quant. Grav. {\bf 18},
5055 (2001).

\bibitem{Ma03b} L. G. Cabral-Rosetti, T. Matos, D. Nunez and R. A.
Sussman,  Class. Quant. Grav. {\bf 19}, 3603 (2002).

\bibitem{Ma03c} J. E.
Lidsey, T. Matos and L. Arturo Urena-Lopez, Phys. Rev. {\bf
D 66}, 023514 (2002).

\bibitem{La03} K. Lake, Phys. Rev. Lett. {\bf 92}, 051101 (2004).

\bibitem{BhKa03}  S. Bharadwaj and S. Kar,  Phys. Rev. {\bf D 68},
023516 (2003).

\bibitem{Sa84}  R. H. Sanders,  Astron. Astrophys. {\bf 136},
L21 (1984).

\bibitem{Sa86} R. H. Sanders, Astron. Astrophys. {\bf 154}, 135
(1986).

\bibitem{Minazzoli:2012md}
O.~Minazzoli and T.~Harko,
Phys. Rev. {\bf D 86}, 087502 (2012).
.


\bibitem{Panpanich:2018cxo}
S.~Panpanich and P.~Burikham,
Phys. Rev. D \textbf{98},  064008 (2018)

\bibitem{Mannheim:2010ti}
P.~D.~Mannheim and J.~G.~O'Brien,
Phys. Rev. Lett. \textbf{106}, 121101 (2011).

\bibitem{Green:2019cqm}
M.~A.~Green and J.~W.~Moffat,
Phys. Dark Univ. \textbf{25},  100323 (2019).

\bibitem{Chardin:2021ahm}
G.~Chardin, Y.~Dubois, G.~Manfredi, B.~Miller and C.~Stahl,
Astron. Astrophys. \textbf{652},  A91 (2021).

\bibitem{Roshan:2021mfc}
M.~Roshan, I.~Banik, N.~Ghafourian, I.~Thies, B.~Famaey, E.~Asencio and P.~Kroupa,
Mon. Not. Roy. Astron. Soc. \textbf{503},  2833 (2021).

\bibitem{Ade:2015xua}
  P.~A.~R.~Ade {\it et al.} [Planck Collaboration],
  Astron.\ Astrophys.\  {\bf 594}, A13 (2016).

\bibitem{Lelli:2016zqa}
F.~Lelli, S.~S.~McGaugh and J.~M.~Schombert,
Astron. J. \textbf{152},  157 (2016).

\bibitem{Sofue:2008wt}
Y.~Sofue, M.~Honma and T.~Omodaka,
Publ. Astron. Soc. Jap. \textbf{61},  227 (2009).

\bibitem{deBlok:2002vgq}
W.~J.~G.~de Blok and A.~Bosma,
Astron. Astrophys. \textbf{385}, 816  (2002).

\bibitem{KuziodeNaray:2007qi}
R.~Kuzio de Naray, S.~S.~McGaugh and W.~J.~G.~de Blok,
Astrophys. J. \textbf{676},  920 (2008).

\bibitem{vizier}
Data from http://vizier.u-strasbg.fr/viz-bin/VizieR.

\bibitem{deBlok:2009sp}
W.~J.~G.~de Blok,
Adv. Astron. \textbf{2010}, 789293  (2010).

\bibitem{BT08} J. Binney and S. Tremaine, Galactic dynamics, Princeton, N. J., Woodstock: Princeton University Press, (2008).

\bibitem{MaLo05} G. A. Mamon and E. L. Lokas, Mon. Not. R. Astron. Soc. {\bf 363}, 705 (2005).

\end{thebibliography}
\end{document}